\documentclass[sigconf, nonacm]{acmart}

\AtBeginDocument{%
  \providecommand\BibTeX{{%
    \normalfont B\kern-0.5em{\scshape i\kern-0.25em b}\kern-0.8em\TeX}}}

\usepackage{enumitem}
\usepackage{tcolorbox}
\usepackage[ruled,linesnumbered,noend]{algorithm2e}
\usepackage{mathrsfs}
\usepackage{makecell}
\usepackage{subcaption}
\usepackage{tablefootnote}
\usepackage{booktabs,tabularx}
\usepackage{bbding}
\usepackage{multirow}
\usepackage{hyperref}
\usepackage{xcolor}
\usepackage{xpatch}
\usepackage{comment}
\let\maketitlesup\maketitle\vspace{3ex}
\xpatchcmd{\maketitlesup}{\@mkteasers}{}{}{}
\xpatchcmd{\maketitlesup}{\@mkabstract}{}{}{}

\setcopyright{none}
\newcommand\vldbdoi{XX.XX/XXX.XX}
\newcommand\vldbpages{XXX-XXX}
\newcommand\vldbvolume{14}
\newcommand\vldbissue{4}
\newcommand\vldbyear{2024}
\newcommand\vldbauthors{\authors}
\newcommand\vldbtitle{\shorttitle} 
\newcommand\vldbavailabilityurl{URL_TO_YOUR_ARTIFACTS}

\newcommand\vldbpagestyle{empty}

\begin{document}
\title{
{Are Joins over LSM-trees Ready: Take RocksDB as an Example} \\ 
 $[$Experiment, Analysis \& Benchmark$]$}

\newcommand{\siqiang}[1]{\textcolor{red}{~~#1}}

\newcommand{\fan}[1]{\textcolor{black}{~~#1}}
\newcommand{\ywp}[1]{\textcolor{orange}{~~#1}}

\newcommand{\newexp}[1]{\textcolor{orange}{~~#1}}

\newcommand{\highlt}[1]{\textcolor{olive}{~~#1}}

\newcommand{\revision}[1]{\textcolor{black}{~~#1}}

\settopmatter{printfolios=false}
\pagestyle{empty}

\setlength{\textfloatsep}{6.0pt plus 0.8pt minus 5.6pt}
\setlength{\floatsep}{6.0pt plus 0.8pt minus 5.2pt}
\setlength{\intextsep}{6.0pt plus 0.8pt minus 5.8pt}
\setlength{\abovedisplayskip}{4.pt plus 0.4pt minus 3.0pt}
\setlength{\belowdisplayskip}{4.pt plus 0.4pt minus 3.0pt}
\tcbset{colframe=white!75!black, boxrule=0.3mm, left=0.4mm, right=0.4mm, top=0.4mm, bottom=0.4mm, arc=0.75mm, after skip=1mm, before skip=2mm}
\setcounter{figure}{0}
\setcounter{table}{0}




\title{
\textcolor{black}{Are Joins over LSM-trees Ready: Take RocksDB as an Example} }




\author{Weiping Yu}
\authornote{Both authors contributed equally to this research.}
\email{weiping001@e.ntu.edu.sg}
\author{Fan Wang}
\authornotemark[1]
\email{FAN008@e.ntu.edu.sg}
\affiliation{%
  \institution{Nanyang Technological University}
}

\author{Xuwei Zhang}
\affiliation{%
  \institution{Nanyang Technological University}
}
\email{zhan0612@e.ntu.edu.sg}

\author{Siqiang Luo}
\authornote{Corresponding Author}
\affiliation{%
  \institution{Nanyang Technological University}
}
\email{siqiang.luo@ntu.edu.sg}
\begin{abstract}
LSM-tree-based data stores are widely adopted in industries for their excellent performance. As data scale increases, disk-based join operations become indispensable yet costly for the database, making the selection of suitable join methods crucial for system optimization. Current LSM-based stores generally adhere to conventional relational database practices and support only a limited number of join methods. However, the LSM-tree delivers distinct read and write efficiency compared to the relational databases, which could accordingly impact the performance of various join methods. Therefore, it is necessary to reconsider the selection of join methods in this context to fully explore the potential of various join algorithms and index designs. 
In this work, we present a systematic study and an exhaustive benchmark for joins over LSM-trees. We define a configuration space for join methods, encompassing various join algorithms, secondary index types, and consistency strategies. We also summarize a theoretical analysis to evaluate the overhead of each join method for an in-depth understanding. Furthermore, we implement all join methods in the configuration space on a unified platform and compare their performance through extensive experiments. Our theoretical and experimental results yield several insights and takeaways tailored to joins in LSM-based stores that aid developers in choosing proper join methods based on their working conditions.

\end{abstract}

\maketitle

\pagestyle{\vldbpagestyle}
\begingroup\small\noindent\raggedright\textbf{PVLDB Reference Format:}\\
\vldbauthors. \vldbtitle. PVLDB, \vldbvolume(\vldbissue): \vldbpages, \vldbyear.\\
\href{https://doi.org/\vldbdoi}{doi:\vldbdoi}
\endgroup
\begingroup
\renewcommand\thefootnote{}\footnote{\noindent
This work is licensed under the Creative Commons BY-NC-ND 4.0 International License. Visit \url{https://creativecommons.org/licenses/by-nc-nd/4.0/} to view a copy of this license. For any use beyond those covered by this license, obtain permission by emailing \href{mailto:info@vldb.org}{info@vldb.org}. Copyright is held by the owner/author(s). Publication rights licensed to the VLDB Endowment. \\
\raggedright Proceedings of the VLDB Endowment, Vol. \vldbvolume, No. \vldbissue\ %
ISSN 2150-8097. \\
\href{https://doi.org/\vldbdoi}{doi:\vldbdoi} \\
}\addtocounter{footnote}{-1}\endgroup

\ifdefempty{\vldbavailabilityurl}{}{
\vspace{.3cm}
\begingroup\small\noindent\raggedright\textbf{PVLDB Artifact Availability:}\\
The source code, data, and/or other artifacts have been made available at \url{https://github.com/we1pingyu/lsmjoin}.
\endgroup
}

\pagestyle{empty}
\setcounter{figure}{0}
\setcounter{table}{0}
\setcounter{section}{0}
\setcounter{equation}{0}
\setcounter{page}{1}

\let\clearpage\relax
\vspace{-2mm}
\section{Introduction}
\label{sec:intro}
Log-structured merge (LSM) trees~\cite{o1996log} based key-value stores have gained significant traction in the industry. 
Notable examples include RocksDB~\cite{rocksdb} at Facebook, LevelDB\cite{google-leveldb} and BigTable~\cite{chang2008bigtable} at Google, HBase~\cite{hbase} and Cassandra~\cite{cassandra} at Apache, X-Engine~\cite{huang2019x} at Alibaba, WiredTiger~\cite{WiredTiger} at MongoDB, and Dynamo~\cite{decandia2007dynamo} at Amazon. These LSM-based key-value stores play crucial roles in various applications such as social media~\cite{armstrong2013linkbench,bortnikov2018accordion}, stream processing~\cite{cao2013logkv,chen2016realtime}, and file systems~\cite{jannen2015betrfs,shetty2013building}.


\vspace{1mm}
\noindent
{\bf Joins over LSM-based stores.}
Many renowned LSM-based stores have implemented join operations~\cite{cassandra,decandia2007dynamo,taft2020cockroachdb,cao2022polardb,rocksdb,matsunobu2020myrocks} as shown in Table ~\ref{table:nosql}. However, they typically rely on insights from 
relational databases to focus on a few join methods. This raises the question: Are these methods still the most appropriate choice for LSM-based stores? 
Our analysis suggests a likely negative answer due to the distinct and complex impact of LSM-trees. Specifically, while LSM-tree enhances updates and lookups efficiency, it requires additional costs to maintain consistency, and the overall impact of LSM-tree has not been meticulously evaluated. Consequently, there may exist substantial space for optimizing the join method selection strategy in LSM-based stores due to two problems.

\begin{figure}[t]
\vspace{-3mm}
\centering
\hspace*{-0.03\linewidth}
\includegraphics[width=1.05\linewidth]{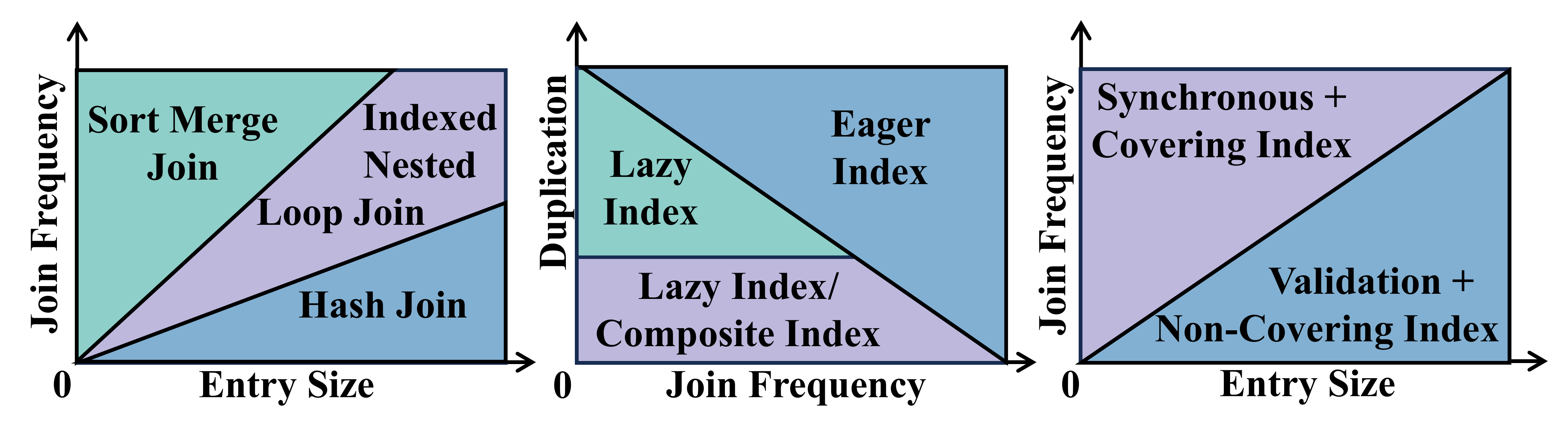}
    \vspace{-5mm}
  \caption{Our benchmark provides helpful guidelines to assist practitioner in selecting proper methods according to their working conditions for join in LSM-based key-value stores.}
\label{fig:insight}
\vspace{-2mm}
\end{figure}

\begin{table*}[t]
\vspace{-3mm}
\footnotesize
\centering
\begin{center}
  \caption{Different join techniques supported in different renowned LSM-based key-value databases.}
  \label{table:nosql}
          \vspace{-3mm}
\begin{tabular}{|l|l|l|l|l|}
\hline
\textbf{NoSQL Storage Systems} & \textbf{Secondary Index} & \textbf{Join Algorithm} & \textbf{Consistency Strategy} & \textbf{Covering Index} \\
\hline
AsterixDB~\cite{altwaijry2014asterixdb} & Embedded Index, Composite Keys & INLJ, SJ, HJ & Synchronous & \checkmark \\
\hline
Cassandra~\cite{cassandra} & Lazy Index & 	$\times$ & Synchronous & $\times$ \\ 
\hline
CockroachDB~\cite{taft2020cockroachdb}, PolarDB-X Engine~\cite{cao2022polardb}
 & Composite Index & INLJ, SJ, HJ & Synchronous & \checkmark \\
 \hline
 HBase~\cite{hbase}, LevelDB~\cite{google-leveldb}, RocksDB~\cite{rocksdb} & $\times$ & 	$\times$ & $\times$ & $\times$ \\ 
\hline
MyRocks~\cite{matsunobu2020myrocks}
 & Composite Index & INLJ
 & Synchronous & \checkmark \\
\hline
\end{tabular}
    \end{center}
        \vspace{-2mm}
\end{table*}

\vspace{1mm}
\noindent{\bf Problem 1: Many join methods remain unexplored in existing LSM-based stores.} As indicated in Table ~\ref{table:nosql}, current LSM-based databases only practice 
 limited join methods. However, the potential of other join techniques has not been fully explored. 
 For instance, most LSM-based stores employ a synchronous consistency strategy instead of a validation one for handling LSM-trees' out-of-place updates. 
 This rationale is supported by the synchronous strategy's advantage in join efficiency, as the validation strategy incurs additional joining overhead.
 However, synchronous strategy simultaneously introduces substantial update overhead which can conversely deteriorate the performance under certain conditions. In such cases, the validation strategy may surprisingly provide better performance. Thus, it is crucial to expand the join design space to include more join methods to enhance the selection criteria.

\vspace{1mm}
\noindent{\bf Problem 2: 
Many influential factors, tied to LSM-tree properties, on join performance have not been fully examined.}  While existing LSM-based stores typically select join algorithms based on selectivity inspired by relational database practices~\cite{wu2023factorjoin,leis2015good,sun13end,wu2021unified}, LSM-tree storage introduces additional performance considerations.
{For instance, the performance of indexed nested loop join (INLJ) maintains robustness, compared to hash join (HJ) and sort-merge join (SJ), for large entry sizes due to the lookup optimization techniques in LSM-tree (e.g. Bloom filters). Hence, though INLJ is generally less preferred in high-selectivity cases, it still can outperform HJ and SJ when handling large entry sizes.}
Therefore, selecting an appropriate join method in LSM-based stores is complex and requires examining more influencing factors. 

To address these problems, we present a systematic study of joins in LSM-based stores, offering an exhaustive benchmark that yields interesting new insights. Our benchmark contributes mainly to the following aspects:

{\bf We identify key join method characteristics and propose an inclusive configuration space covering both existing and potential new combinations.} This configuration space includes four primitive join components, join algorithm (i.e. indexed nested loop join, sort-merge join, and hash join), secondary index (i.e. eager index, lazy index, and composite index), consistency strategy (i.e. synchronous strategy and validation strategy), and covering index (i.e. covering and non-covering). It allows us to describe existing join methods and discover novel ones, such as the indexed nested loop join combined with a non-covering eager index and validation strategy. Such an integrated study allows us to explore the join method overlooked by existing literature, which typically concentrates on certain aspects of the entire space~\cite{qader2018comparative,luo2019efficient,wang2023revisiting}.

{\bf We tailor the theoretical analysis to the joins over LSM-tree including 3 join algorithms across 12 scenarios and 6 index designs, many of which have not been previously analyzed in existing works.}
Different from existing join analysis, our assessment incorporates the unique characteristics of LSM-trees to evaluate the overhead of join methods. Additionally, we analyze the integrated cost of each join method in our configuration space, considering its distinct join algorithm, secondary index type, consistency strategy, and covering index. Many combinations, such as the integration of eager index designs with validation strategies, have not been explored in prior research. This thorough analysis not only leads to an in-depth understanding of the features of each join method but also guides the design of our experiments.

{\bf We implement all 29 join methods within our configuration space on a unified platform and examine them under diverse working conditions to derive guidelines for join method selection in LSM-based stores.} 
We examine 10 factors related to workload and LSM-tree configuration to assess their impact on the performance of diverse join methods.
Excitingly, this leads to several useful guidelines as Figure ~\ref{fig:insight} illustrates. 
In the {\it Movie} dataset with moderate selectivity, SJ is the preferred method based solely on selectivity. However, as entry size increases to 4096 bytes, INLJ outperforms both SJ and HJ by 70\%, making it the best choice.
Notably, some differ from traditional relational databases and other data storage systems like those based on B+ trees. For example, while selectivity is typically the main criterion in relational databases, our results show that join frequency and entry size also play crucial roles. In the {\it Movie} dataset with moderate selectivity, SJ is the preferred method based solely on selectivity. However, as entry size increases to 4096 bytes, INLJ outperforms both SJ and HJ by 70\%, making it the best choice.
Additionally, some overlooked secondary index designs can offer competitive performance. For instance, while the eager index is usually avoided due to its high construction cost, it excels in frequent join workloads. In the {\it User} dataset, INLJ with a composite index performs better than the eager index when joins occur less than once per 10 million updates. However, when join frequency increases to 32, the eager index outperforms the composite index by over 30\% in latency.

The remainder of this paper is organized as follows: Section ~\ref{sec:pre} provides preliminary knowledge about LSM-trees and joins in LSM-based stores; Section ~\ref{sec:configuration} introduces our configuration space; the experimental results and discussion are elucidated in Section ~\ref{sec:evaluation}; Section ~\ref{sec:highlight} summarizes the important insights and takeaways to enhance our understanding of this topic, as well as points out some directions to guide future research.

\begin{table}[t]
\vspace{-1mm}
\footnotesize
\centering
\begin{center}
  \caption{Notations used in this paper. Subscripts denote associated tables (e.g., $L_R$: the number of levels in right data table LSM-tree, $L_{R'}$: the number of levels in right index LSM-tree).}
      \renewcommand{\arraystretch}{1.05}
          \vspace{-3mm}
\begin{tabular}{l|p{5.2cm}|l}
\toprule
\textbf{Term} & \textbf{Definition}     & \textbf{Unit} \\ \hline
$R, S$    & Right and left data tables&        \\ 
$R', S'$    & Index tables for right and left data tables &  \\
$L$  & Number of levels in an LSM-trees &  levels \\
$N$  & Number of entries stored in an LSM-tree &  entries \\
$e$  & Size of entries of LSM-tree  &  bytes \\
$B$  & Size of data block &  bytes \\ 
$p$  &False positive rate of Bloom filters  &         \\ 
$\epsilon$  & Ratio between the number of matched entries and total entries in the LSM-tree  &         \\ 
$d$  & Average duplication frequency of join attribute value in a sequence of updates  &   entries  \\ 
\bottomrule
\end{tabular}
\label{table:notation}
    \end{center}
    \vspace{-2mm}
\end{table}

\begin{figure*}[t]
\vspace{-3mm}
\centering
  \includegraphics[width=0.98\linewidth]{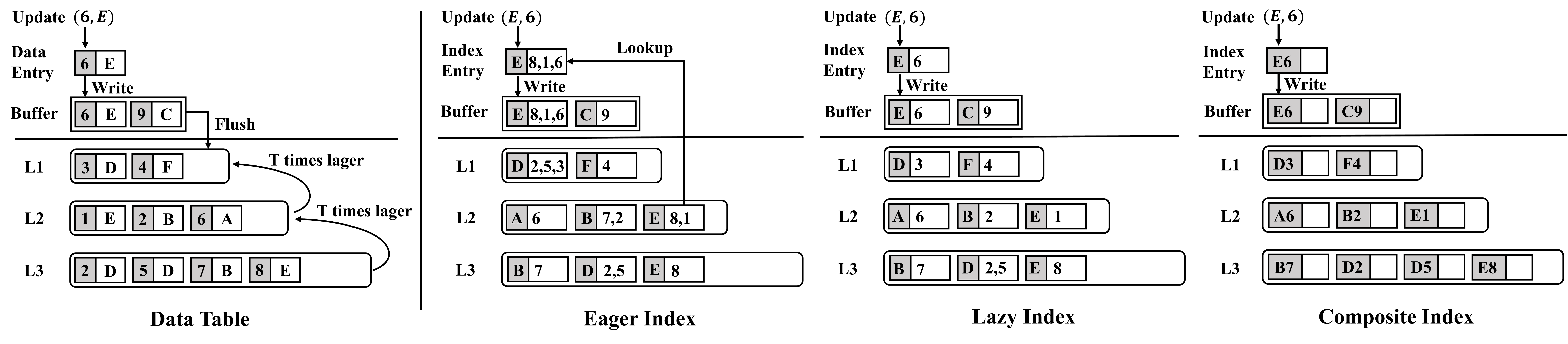}
  \vspace{-3mm}
  \caption{An illustration of LSM-trees for a data table and the corresponding index table, constructed using various secondary index types. In this figure, dark blocks represent the keys of entries, while white blocks represent the values.}
\label{fig:sec_index}
\vspace{-3mm}
\end{figure*}

\vspace{-1mm}
\section{Background}
\label{sec:pre}
This section discusses background knowledge about joins over LSM-based databases. Frequently used notations are listed in Table~\ref{table:notation}.

\vspace{-1mm}
\subsection{Log-Structured Merge Trees}
An LSM-tree organizes data using an in-memory write buffer and multiple on-disk levels with capacities increasing exponentially by a size ratio $T$. To store $N$ entries of size $e$, an LSM-tree should incorporate $L = \log_T \frac{N \cdot e}{M}$ levels, where $M$ is the write buffer size. LSM-tree supports mainly three types of operations as follows. 

\noindent\textbf{Updates.} LSM-tree uses an out-of-place update strategy where key-value pairs, or {\it entries}, are initially stored in the write buffer. When this buffer is full, the entries are flushed to disk, progressing through the LSM-tree levels as shown in Figure \ref{fig:sec_index}. In the worst case, an entry reaches the largest level after $L \cdot T$ compaction processes, resulting in an update cost of $O(L \cdot T \cdot \frac{e}{B})$\cite{dayan2018dostoevsky,dayan2017monkey}.  

\noindent\textbf{Point Lookups.} 
A point lookup in an LSM-tree searches for an {\it entry} using a given {\it key}. It checks through all levels, returning the result once the matched entry is found. To speed up this process, Bloom filters~\cite{bloom1970space} are employed to indicate the presence of the matched entry with false positive rate $p$. Thus the lookup cost is $O(L \cdot p)$ if the entry is absent, or $O(L \cdot p + \lceil \frac{e}{B} \rceil)$ if present.

\noindent\textbf{Range Lookups.} 
A range lookup retrieves entries within a key range in an LSM-tree. It incurs $O(L)$ I/O cost to seek qualified entries across all levels~\cite{dayan2018dostoevsky,dayan2017monkey}, and requires an additional $O(\frac{d \cdot e}{B})$ cost for entry retrieval, where $d$ is the number of matched entries.


\vspace{-2mm}
\subsection{Joins in LSM-tree based databases}
\label{subsec:bg_join}

Joins in LSM-tree based stores typically fall into two scenarios based on the alignment of the join attribute with the primary keys of the involved data tables. We examine an elementary case involving two {\it data tables}, where the left and right tables are denoted as \( R \) and \( S \). The data tables are stored in separate LSM-trees with their primary keys as the keys in LSM-tree entries. We represent these tables as \( R(p_r, n_r) \) and \( S(p_s, n_s) \), where $p$ is the primary key and $n$ represents non-primary attributes.
\revision{
Hence, there are two possible scenarios for the data tables in join operation. 
In one scenario, the join attribute is the primary key of the data table, thus entries with specific join attributes can be accessed directly via the LSM key (referred to as primary index) without additional overhead. 
The other scenario is more complex where the join attribute is a non-primary attribute. In this case, entries are difficult to identify by the join attribute since non-primary attributes are stored in the values of the LSM-tree entries. Therefore, secondary indexes can be introduced to identify entries by join attributes and then benefiting join algorithms involving lookups (e.g., indexed nested loop joins), and the corresponding {\it index tables} are denoted as \( R' \) and \( S' \), respectively. 
Meanwhile, this introduces a tradeoff between potential join performance gains and additional index construction overhead. 
Our benchmark includes both scenarios and various secondary index types to provide a comprehensive analysis.}

\section{Consolidated join configuration space for LSM-tree}
\label{sec:configuration}

{
\begin{algorithm}
\color{black}
\footnotesize
\DontPrintSemicolon
\KwIn{Data table $R$, index table $R'$, update data $(p_i, n_i)$ \;}
Lookup $p_i$ in $R$ for the corresponding secondary key $n_x$\;
\uIf{$n_x$ == $NULL$}{
    Insert $(p_i, n_i)$ to $R'$\;
}
\Else{
    Lookup $n_x$ in $R'$ for the corresponding primary key $p_x$ \;
    Delete $(n_x, p_x)$ in $R'$\;
    Insert $(n_i, p_i)$ to $R'$\;
}
\caption{Update with Synchronous}
\label{alg:syn}
\vspace{-1mm}
\end{algorithm}
}

\begin{table*}[t]
\footnotesize
\centering
\begin{center}
  \caption{The theoretical analysis of various secondary indexes and consistency strategies. In this table, the point lookup cost refers to the I/O cost of finding index entries associated with a certain join attribute value. Moreover, it is worth mentioning that we only present the cost for updating the index table given the update cost of the data table is identical for each method. }
  \label{tab: sec_index}
      \renewcommand{\arraystretch}{1.5}
          \vspace{-3mm}
\begin{tabular}{|l|l|l|l|l|l|}
\hline
{\bf Index} 
&\textbf{\begin{tabular}[c]{@{}c@{}}Consistency\\strategy\end{tabular}} 
&\textbf{\begin{tabular}[c]{@{}c@{}}Index\\Type\end{tabular}} 
&\textbf{\begin{tabular}[c]{@{}c@{}}Empty Point\\Lookup ($Z_0$)\end{tabular}}
&\multicolumn{1}{c|}{\bf Non-empty Point Lookup ($Z_1$)}
&\multicolumn{1}{c|}{\bf Update (U)}
\\ \hline

S-Eager & \multirow{3}{*}{Synchronous} & Eager Index 
& $O(L' \cdot p)$
& $O(L' \cdot p + \lceil \frac{e'}{B} \rceil)$ 
& $O(L \cdot p + \lceil \frac{e}{B} \rceil) + O(L' \cdot p + \lceil \frac{e'}{B}\rceil) + O(L' \cdot T \cdot \frac{e'}{B}) $
\\ \cline{1-1} \cline{3-6}

S-Lazy &  & Lazy Index 
& $O(L' \cdot p)$
& $O(L' \cdot \lceil \frac{e'}{B} \rceil)$ 
& $O(L\cdot p + \lceil \frac{e}{B} \rceil) + O(L'\cdot T \cdot \frac{e'}{B})$    
\\ \cline{1-1} \cline{3-6}

S-Comp &  & Composite Keys 
& $O(L' \cdot p)$
& $O(L' + d \cdot \frac{e'}{B})$ 
& $O(L\cdot p + \lceil \frac{e}{B} \rceil) + O(L'\cdot T \cdot \frac{e'}{B})$    
\\ \hline

V-Eager & \multirow{3}{*}{Validation} & Eager Index 
& $O(L' \cdot p)$
& $O(d \cdot (L \cdot p + \lceil \frac{e}{B} \rceil)) + O(L' \cdot p + \lceil \frac{e'}{B}\rceil)$  
& $O(L' \cdot p + \lceil \frac{e}{B} \rceil)) + O(L'\cdot T \cdot \frac{e'}{B})$   
\\ \cline{1-1} \cline{3-6}

V-Lazy &  & Lazy Index 
& $O(L' \cdot p)$
& $O(d \cdot (L\cdot p + \lceil \frac{e}{B} \rceil)) + O(L' \cdot \lceil \frac{e'}{B} \rceil)$ 
& $O(L'\cdot T \cdot \frac{e'}{B})$   
\\ \cline{1-1} \cline{3-6}

V-Comp &  & Composite Keys 
& $O(L' \cdot p)$
& $O(d \cdot (L\cdot p + \lceil \frac{e}{B} \rceil)) + O(L' + d \cdot \frac{e'}{B})$ 
& $O(L'\cdot T \cdot \frac{e'}{B})$    
\\ \hline

\end{tabular}
    \end{center}
\end{table*}

Our investigation reveals that existing renowned LSM-based stores support only a limited number of join methods, leaving many potentially high-performing methods unexplored. This gap indicates a lack of systematic study of join methods in LSM-based stores, for which a comprehensive configuration space encompassing a wide range of instances is necessary.
To fill this gap we are the first to propose an inclusive configuration space tailored for joins over LSM-trees comprising various join algorithms, secondary indexes, consistency strategies, and covering indexes. This configuration encompasses all the existing join methods and many combinations that have never been discussed before. Additionally, we provide a theoretical analysis tailored to LSM-trees for each join method within the configuration space, which encourages a more comprehensive and thorough understanding of joins over LSM-trees.

This section provides a general introduction to and theoretical analysis of various index designs (i.e., index types, consistency strategies, and covering indexes) and join algorithms.

\subsection{Secondary index types analysis}
\label{subsec:index}
In LSM-based key-value stores, secondary indexes present a novel trade-off among join performance, update overhead, and space consumption that requires detailed analysis.
\textcolor{black}{We consider three secondary index types~\cite{qader2018comparative,luo2019efficient}: \textit{Eager Index} (Eager), \textit{Lazy Index} (Lazy), and \textit{Composite Index} (Comp).} Our analysis in Table ~\ref{tab: sec_index} quantitatively evaluates their distinct performance to guide the index selection for specific conditions.
\noindent
{\bf Eager Index.} 
Eager Index uses join attributes as keys and stores associated primary keys and attributes in a posting list as the value of the index entry. This structure links multiple entries that share the same join attribute value. Its update process requires an extra point lookup to retrieve the existing index entry with the same join attribute, and then updates this entry before inserting it into the index LSM-tree. 
It incurs $O(L \cdot p + \lceil \frac{e}{B} \rceil)$ I/O cost for retrieval and $O(L \cdot T \cdot \frac{e}{B})$ for updates, as exemplified in Figure ~\ref{fig:sec_index}. The cost of data extraction via point lookups is $O(L\cdot p)$ for empty lookups and $O(L\cdot p + \lceil \frac{e}{B} \rceil)$ for non-empty ones. The low false positive rate of Bloom filters suggests a lower cost for point lookups, potentially favoring eager indexes and indexed nested loop joins for certain workloads, as discussed in the evaluation section.

\noindent
{\bf Lazy Index.} Lazy Index also uses posting lists to store associated primary keys within an index entry but defers their merging. Unlike Eager Index, it merely appends a new entry to the write buffer, with index entries gradually merging during compaction at a cost of $O(L \cdot T \cdot \frac{e}{B})$. As a result, entries with the same key may exist concurrently at different LSM-tree levels, requiring a modified lookup process that examines all levels to aggregate all relevant entries. This results in a lookup cost of $O(L \cdot \lceil \frac{e}{B} \rceil)$. For instance, as Figure ~\ref{fig:sec_index} shows, retrieving entries with join attribute ${E}$ requires extracting index entries from multiple levels to compile the complete posting list $\{6,1,8\}$. Lazy Index provides better update efficiency but incurs higher lookup overhead, suiting workloads requiring rapid index updates but tolerating infrequent point lookups.

\noindent
{\bf Composite Index.} The secondary index with a composite key stores tuples with the same join attribute value in separate entries, using a concatenated index key of the primary key and join attribute. This setup enhances lookups with prefix filters, where the join attribute serves as the prefix. For example, as illustrated in Figure ~\ref{fig:sec_index}, n update involving the primary key $\{6\}$ and join attribute $\{E\}$ generates an index key $\{E6\}$, incurring an update cost of $O(L \cdot T \cdot \frac{e}{B})$. Point lookups require examining all LSM-tree levels due to the potential distribution of entries, with a cost of $O(L + d \cdot \frac{e}{B})$, where $d$ is the number of matched entries. Although Composite Index generally offers robust update and lookup performance, its practicality may suffer due to increased space requirements and more compaction cost, particularly with more duplicated join attributes.


\noindent 
\textcolor{black}{{\bf Runing example:}
A data table with 10 million 64 bytes tuples introduces a three-level LSM-tree if the write buffer is 16 MB and size ratio is 10. When the Bloom filter is set to 10 bits per key, according to Table ~\ref{tab: sec_index}, the searching cost $(L' \cdot p)$ of Eager is neglectable compared to the other two types, indicating faster lookup performance. Meanwhile, the update cost of Eager is much higher since the writing cost $(L \cdot T \cdot \frac{e}{B})$ is less than 1 while the additional cost $(L \cdot p + \lceil \frac{e}{B} \rceil)$ is greater than 1. Hence Eager benefits query-intensive workloads like indexed nested loop join with frequent join.}



\begin{figure}[t!]
\centering
  \includegraphics[width=0.99\linewidth]{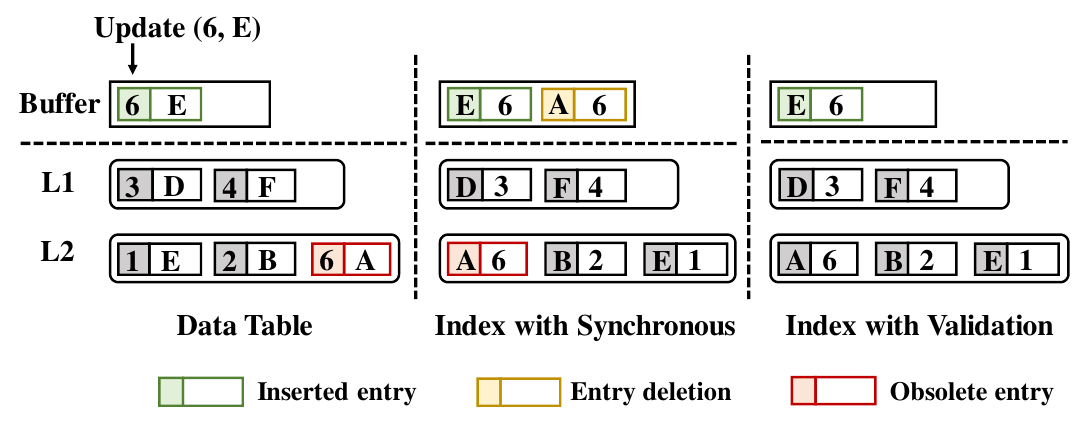} 
  \vspace{-2mm}
  \caption{\textcolor{black}{Synchronous incurs extra update cost, Validation cannot remove invalid entries, adding query overhead.}}
\label{fig:consistency}
\end{figure}

\subsection{Consistency strategy analysis}
\label{subsec:consistency}

\begin{table*}[t]
\vspace{-1mm}
\footnotesize
\centering
\begin{center}
  \caption{\revision{Theoretical analysis of join cost, and $S$ and $R$ represent the inner and outer table for INLJ, respectively.}
  }
  \label{tab: join_cost}
      \renewcommand{\arraystretch}{1.5}
  \vspace{-3mm}
\begin{tabular}{|l|c|l|l|l|l|}
\hline
\textbf{Method}
& \textbf{Join algorithm}
&\textbf{Data table S} 
&\textbf{Data table R} 
&\textbf{Space complexity}
&\textbf{Join cost}\\ \hline

INLJ-P & \multirow{5}{*}{\begin{tabular}[c]{@{}c@{}}Indexed Nested\\Loop Join\end{tabular}} & Primary index & Regular column 
& $D(R)+D(S)$ 
& $(N_R \cdot \frac{e_R}{B}) + N_R((1 -\epsilon_{R}) Z_0(S) + \epsilon_{R} \cdot Z_1(S))$ 
\\ \cline{1-1} \cline{3-6}

INLJ-PS & & Primary index & Secondary index 
& $D(R)+D(S)+D(R')$ 
& $(N_{R'} \cdot \frac{e_{R'}}{B}) + N_R((1 -\epsilon_{R}) Z_0(S) + \epsilon_{R} \cdot Z_1(S))$ 
\\ \cline{1-1} \cline{3-6}

INLJ-N & & Regular column & Regular column 
& $D(R)+D(S)$ 
& $(N_R \cdot \frac{e_R}{B}) + N_R (N_S \cdot \frac{e_S}{B})$
\\ \cline{1-1} \cline{3-6}

INLJ-NS & & Secondary index & Regular column 
& $D(R)+D(S)+D(S)$ 
& $(N_R \cdot \frac{e_R}{B}) + N_R((1 -\epsilon_{R}) Z_0(S') + \epsilon_{R} \cdot Z_1(S'))$
\\ \cline{1-1} \cline{3-6}

INLJ-SS & & Secondary index & Secondary index 
& $D(R)+D(S)+D(R')+D(S')$ 
& $(N_{R'} \cdot \frac{e_{R'}}{B}) + N_R((1 -\epsilon_{R}) Z_0(S') + \epsilon_{R} \cdot Z_1(S'))$ 
\\ \hline

SJ-P &\multirow{5}{*}{\begin{tabular}[c]{@{}c@{}}Sort-merge\\Join\end{tabular}} & Primary index & Regular column 
& $2D(R)+2D(S)$ 
& $(N_{S} \cdot \frac{e_S}{B}) + 5(N_R \cdot \frac{e_R}{B})$ 
\\ \cline{1-1} \cline{3-6}

SJ-PS & & Primary index & Secondary index 
& $D(R)+2D(S)+D(R')$  
& $(N_{S} \cdot \frac{e_S}{B}) +(N_{R'} \cdot \frac{e_{R'}}{B})$
\\ \cline{1-1} \cline{3-6}

SJ-N & & Regular column & Regular column 
& $2D(R)+2D(S)$  
& $5(N_S \cdot \frac{e_S}{B}) + 5(N_R \cdot \frac{e_R}{B})$
\\ \cline{1-1} \cline{3-6}

SJ-NS & & Secondary index & Regular column 
& $2D(R)+D(S)+D(S')$  
& $(N_{S'} \cdot \frac{e_{S'}}{B}) + 5(N_R \cdot \frac{e_R}{B})$
\\ \cline{1-1} \cline{3-6}

SJ-SS & & Secondary index & Secondary index 
& $D(R)+D(S)+D(R')+D(S')$  
& $(N_{S'} \cdot \frac{e_{S'}}{B}) +(N_{R'} \cdot \frac{e_{R'}}{B})$
\\ \hline

HJ-P &\multirow{2}{*}{Hash Join} & Primary index & Regular column 
& $2D(R)+2D(S)$ 
& $3(N_S \cdot \frac{e_S}{B}) + 3(N_R \cdot \frac{e_R}{B})$ 
\\ \cline{1-1} \cline{3-6}

HJ-N & & Regular column & Regular column 
& $2D(R)+2D(S)$ 
& $3(N_S \cdot \frac{e_S}{B}) + 3(N_R \cdot \frac{e_R}{B})$
\\ \hline

\end{tabular}
    \end{center}
    \vspace{-2mm}
\end{table*}

Due to the out-of-place update mechanism, obsolete data entries in LSM-based systems may not be promptly removed, requiring consistent maintenance between data and index tables to ensure accurate join results that are achieved via typical two methods. 

{
\begin{algorithm}
\color{black}
\footnotesize
\DontPrintSemicolon
\vspace{-1mm}
\KwIn{Data table $R$, index table $R'$, queried secondary key $n_i$ \;
  }
Lookup $n_i$ in $R'$ for the corresponding primary key $p_x$\;
\uIf{$p_x$ is $NULL$}{
    \Return $NULL$  \;
}
\Else{
    Lookup $p_x$ in $R$ for the corresponding secondary key $n_x$ \;
    \uIf{$n_x$ == $n_i$}{
    \Return $(n_i, p_i)$  \;
    }
    \Else{
     \Return $NULL$
    }
}
\caption{Query with Validation Strategy}
\label{alg:val}
\vspace{-1mm}
\end{algorithm}
}

\noindent {\bf Synchronous.} Synchronous immediately checks validity upon updates to maintain consistency. This strategy improves lookup performance but increases update costs due to additional synchronization. The update process involves verifying data entry existence and updating the index. Overall update costs for different secondary indexes are detailed in Table ~\ref{tab: sec_index}.
\noindent {\bf Validation.} 
Validation in LSM-style secondary indexes ensures join result correctness by checking the data table during joins. Hence, it simply need to insert new entry to update the index LSM-tree without removing obsolete data.
Accordingly, the joining cost of Validation includes both querying the index entry and verifying all associated primary keys, expressed
as $O(d \cdot (L\cdot p + \lceil \frac{e}{B} \rceil)) + O(L' + d \cdot \frac{e'}{B})$ 
in Table ~\ref{tab: sec_index}, where $d$ is the number of primary key duplicates per index entry. Validation therefore incurs lower costs during index building but higher costs during joining.



\noindent {\bf Covering Index.} 
A covering index includes the primary key and all tuple attributes, except the join attribute, within the index entry's value. Conversely, a non-covering index only stores the secondary and primary key, resulting in smaller entry sizes which delivers improved update and degraded lookup performance. The choice between covering and non-covering index is closely tied to the consistency strategy. Covering indexes complement Synchronous methods by enhancing lookup performance without querying data LSM-trees. Validation pairs well with non-covering indexes, as both require data entry retrieval. 



\vspace{-2mm}
\subsection{Join algorithm analysis}
\label{subsec:join}

\textcolor{black}{In this section, we integrate the previously discussed index designs into three join algorithms: {\it indexed nested loop join} (INLJ), {\it sort-merge join} (SJ), and {\it hash-based join} (HJ), in LSM-based key-value stores.} A join method is the combination of specific join algorithm, secondary index type, and consistency strategy. 
\noindent 
{\bf Indexed Nested Loop Join.}
\revision{INLJ is a fundamental join that scans the left (outer) data table $R$ and searches the right (inner) data table $S$ for matching join attributes.} This method's overhead includes scanning and searching costs. Scanning each entry in $R$'s LSM-tree incurs a constant cost of $N_R \cdot \frac{e_R}{B}$. Without an index, searching requires scanning the entire right LSM-tree for each $R$ entry, costing $N_R \cdot (N_S \cdot \frac{e_S}{B})$, which is unacceptable. However, if $S$ is indexed, either by its primary key or a separate index LSM-tree, 
the cost is drastically reduced by the point lookup operation.
As detailed in Section~\ref{subsec:index}, point lookup costs depend on whether the search finds a match. Assuming a match ratio $\epsilon_R$ in $R$, the search cost is \revision{$(N_R \cdot \frac{e_R}{B}) + N_R((1 -\epsilon_{R}) Z_0 + \epsilon_{R} \cdot Z_1)$}, where $Z_0$ and $Z_1$ represent the costs of empty and non-empty lookups shown in Table ~\ref{tab: sec_index}. The overall join costs for various scenarios are further detailed in Table ~\ref{tab: join_cost}.


\noindent 
{\bf Sort-Merge Join.}
SJ identifies matching tuples by sorting join attributes from the left and right data tables. To this end, entries are scanned from the LSM-trees, sorted into multiple runs, and written back to disk. Note that single-run sorting is not feasible due to memory limits. Subsequently, these runs are processed by a k-way merge~\cite{bentley1984programming} and then saved to disk which is finally read out to execute the join. Hence the join cost sums up to $5(N \cdot \frac{e}{B})$.With available indexes, sorting and merging steps can be skipped since entries can be retrieved sequentially by join attributes. As shown in Table~\ref{tab: join_cost}, SJ performs best when both tables are indexed, which our experimental results confirm.


\noindent 
{\bf Hash Join.}
HJ groups tuples with matching attributes from the left and right tables into the same buckets using join attributes. For left table $R$, it scans all entries, assigns them to buckets based on hash values, and stores them on disk. During the join, these buckets are retrieved, resulting in a total cost of $3(N_R \cdot \frac{e_R}{B})$. Unlike SJ, HJ's efficiency is not influenced by the presence of indexes. As shown in Table ~\ref{tab: join_cost}, HJ generally has lower join overhead than SJ, but SJ performs better when indexes are used. Further analysis of both algorithms' performance is provided in later sections.

\noindent 
\textcolor{black}{{\bf Runing example:}
Consider join of two data tables (*-P), both consisting of 10 million 64 bytes tuples, has matching rate of 1\%. The data block size is 4KB, and Bloom filter uses 10 bits per key. According to Table ~\ref{tab: join_cost}, SJ and HJ exhibit comparable performance, while INLJ outperforms them due to efficient query operations (small $Z_0$ and $Z_1$). When $e_R$, the entry size of table R, decreases from 64 to 4 bytes, SJ’s join cost decreases substantially until surpassing INLJ, whose performance remains stable due to the ceiling component in $Z_1$. Conversely, as $e_R$ increases, HJ can outperform SJ.}

\begin{figure}[t!]
\centering
  \includegraphics[width=0.8\linewidth]{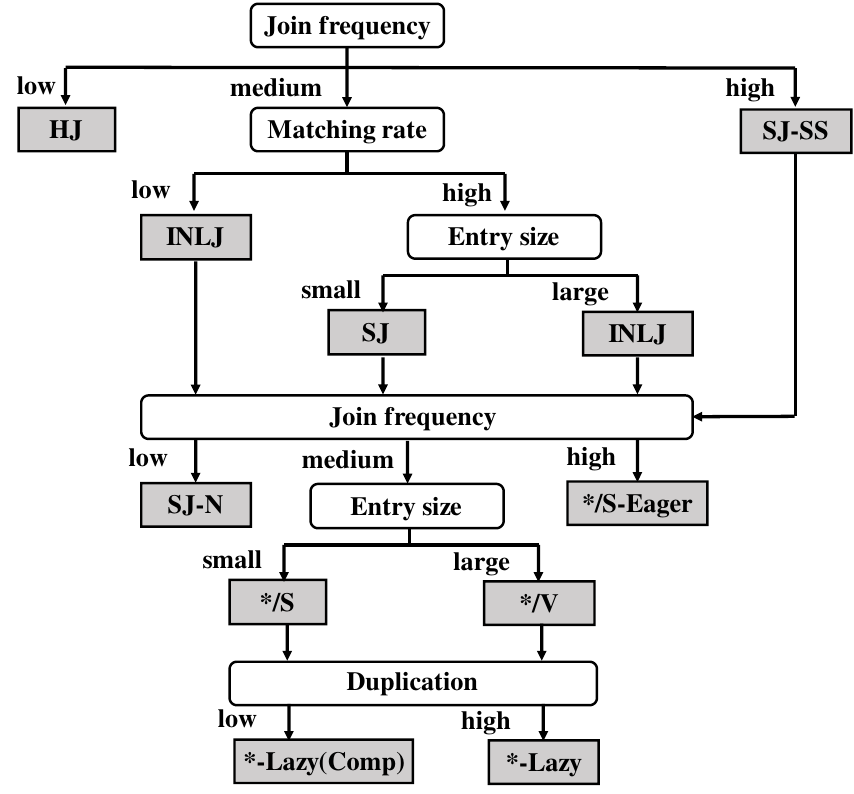}  
  \vspace{-3mm}
  \caption{\textcolor{black}{A sketch of a decision graph according to the theoretical analysis results in Table ~\ref{tab: sec_index} and Table ~\ref{tab: join_cost}.}}
\label{fig:decision_graph}
    \vspace{-2mm}
\end{figure}
\section{Evaluating joins over LSM-trees} 
\label{sec:evaluation}


We experimentally evaluate the influence of various factors on the performance of diverse join methods to summarize practical guidelines for choosing the most suitable methods under specific working conditions. To this end, we identify several important variants introduced in the theoretical analysis as presented in Tables~\ref{tab: sec_index} and~\ref{tab: join_cost}, which can be classified into two categories: {\bf join algorithm related factors} and {\bf secondary index related factors}.
The following parts examine each of these factors individually.
\noindent\textbf{Hardware.}
All experiments are conducted on a server with an Intel Xeon Gold 6326 processors, 256GB DDR4 main memory, 1TB NVMe SSD, a default page size of 4 KB, and running 64-bit Ubuntu 20.04.4 LTS on an ext4 partition.


\noindent\textbf{Implementation.} 
We implement all 29 join methods within our proposed configuration space on a unified platform RocksDB~\cite{rocksdb}, a prevalent LSM-based storage. While RocksDB does not inherently support joins, this required us to design custom workflows for various join algorithms and index types, many of which were developed from scratch for LSM-based stores. For instance, in the Eager and Lazy index designs, long posting lists, particularly with skewed or duplicated data, introduce significant overhead. To mitigate this and improve efficiency, we applied optimized C++ techniques for string management~\cite{meyers2001effective}.
The LSM-tree construction for both data and index tables follows RocksDB's default settings, with a block size of 4096 bytes, a 16MB write buffer, a size ratio of 5 to introduce more levels, and no block cache.
Our implementation of sort-merge join using a streamlined approach, setting the total sorted segment size to 16MB. For hash join, we integrated BKDR hash function~\cite{partow2013general} with Grace Hash Join to reduce hash collisions, using 16MB memory for partitioning and probing.



\begin{table}[t]
\footnotesize
\setlength{\tabcolsep}{4.5pt}
\caption{Feature of Data Distribution in Benchmark Datasets.}
  \label{tab:dataset}
    \vspace{-2mm}
\begin{tabular}{|ll|ll|ll|ll|}
\hline
\multicolumn{2}{|l|}{\multirow{2}{*}{Dataset}}                      & \multicolumn{2}{l|}{Duplication}         & \multicolumn{2}{l|}{Matching}              & \multicolumn{2}{l|}{Skewness}                                                    \\ \cline{3-8} 
\multicolumn{2}{|l|}{}                                                                                       & \multicolumn{1}{l|}{$d_r$}     & $d_s$     & \multicolumn{1}{l|}{$\epsilon_r$} & $\epsilon_s$ & \multicolumn{1}{l|}{$\theta_r$} & $\theta_s$ \\ \hline
\multicolumn{1}{|l|}{\multirow{4}{*}{\begin{tabular}[c]{@{}l@{}}Real\\ Database\end{tabular}}}   
& \it{Movie} & \multicolumn{1}{l|}{4.1}      & 5.6      & \multicolumn{1}{l|}{0.72}       & 0.99       & \multicolumn{1}{l|}{0.89}                & 0.85                \\ \cline{2-8} 
\multicolumn{1}{|l|}{}                                                                               & \it{User}  & \multicolumn{1}{l|}{4.2}      & 3.9      & \multicolumn{1}{l|}{0.42}       & 0.40       & \multicolumn{1}{l|}{0.93}                & 0.92               \\ \cline{2-8}
\multicolumn{1}{|l|}{} 
& \it{Face}  & \multicolumn{1}{l|}{1.0}      & 1.0      & \multicolumn{1}{l|}{1}          & 1& \multicolumn{1}{l|}{0}                            & 0                            \\ \cline{2-8} 
\multicolumn{1}{|l|}{}                                                                               & \it{Wiki}  & \multicolumn{1}{l|}{1.1}      & 1.1      & \multicolumn{1}{l|}{1}          & 1          & \multicolumn{1}{l|}{0.20}                & 0.20                \\ \hline
\multicolumn{1}{|l|}{\multirow{2}{*}{\begin{tabular}[c]{@{}l@{}}Synthetic \\ Datasets\end{tabular}}} & \it{Unif}  & \multicolumn{1}{l|}{4.0} & 4.0 & \multicolumn{1}{l|}{0.8}    & 0.8    & \multicolumn{1}{l|}{0}                            & 0                            \\ \cline{2-8} 
\multicolumn{1}{|l|}{}                                                                               & \it{Zipf}  & \multicolumn{1}{l|}{2.6} & 2.6 & \multicolumn{1}{l|}{1}    & 1    & \multicolumn{1}{l|}{0.5}                      & 0.5                        \\ \hline
\end{tabular}

  \vspace{-2mm}
\end{table}





\begin{figure*}[t]
\vspace{-3mm}
  \centering
  \hspace{-2mm}
    \begin{subfigure}[b]{0.37\textwidth}
            \centering
  \includegraphics[width=\linewidth]{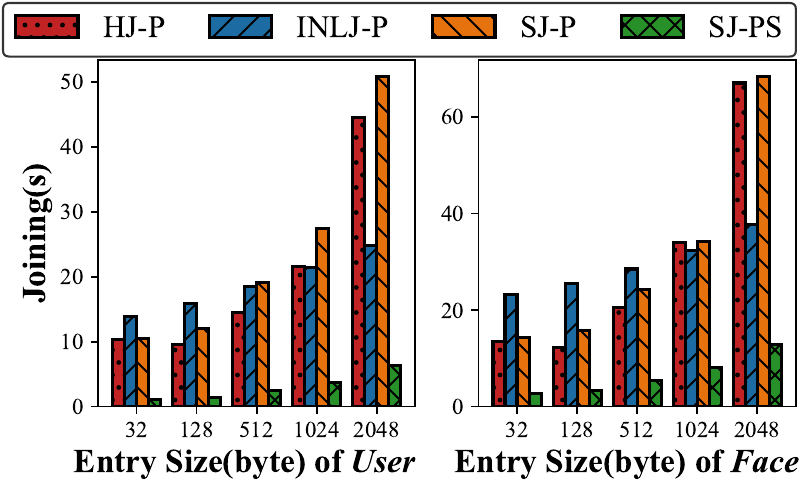}  
                \vspace{-5mm} 
  \caption{Entry Size}
\label{fig:4.2}
  \end{subfigure}
    \hspace{1mm}
   \begin{subfigure}[b]{0.366\textwidth}
           \centering
  \includegraphics[width=\linewidth]{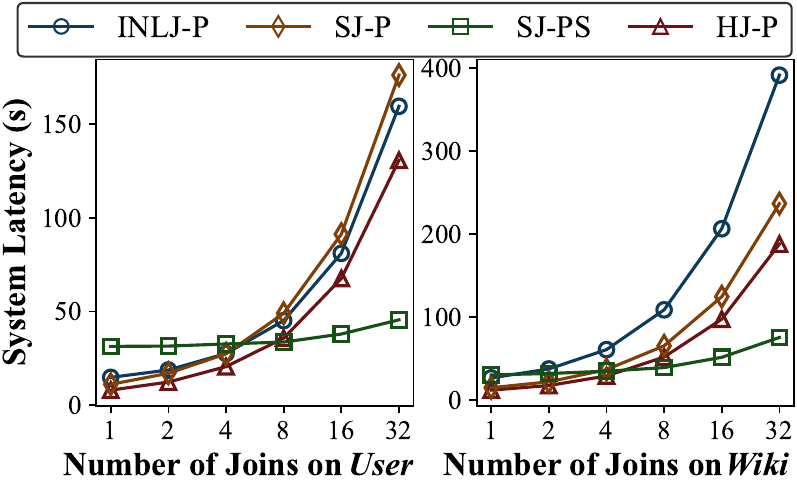}
                \vspace{-5mm} 
  \caption{Join Frequency}
\label{fig:4.3}
  \end{subfigure}
      \hspace{1mm}
  \begin{subfigure}[b]{0.23\textwidth}
           \centering
  \includegraphics[width=\linewidth]{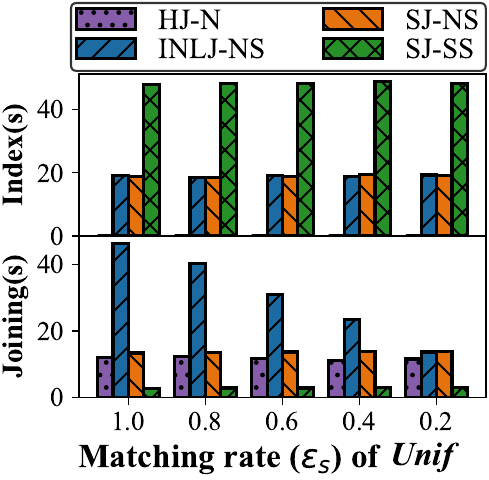}  
  \vspace{-5mm}
  \caption{Matching Rate.}
\label{fig:5.5}
  \end{subfigure} 
  \hspace{-3mm}
      \vspace{-3mm}
  \caption{ Join algorithm performance varies greatly with changes in entry size, join frequency, and matching rate. In (a), SJ-PS incurs additional index building latency (in seconds) as follows: 29, 43, 278, 627, 1162 ({\it User}); 25, 43, 239, 597, 1149 ({\it Face}).
  }
  \label{fig:joinsel}
      \vspace{-3mm}
\end{figure*}

\noindent\textbf{Baselines.}
We examine all join methods in our configuration space each specified by particular join algorithm (i.e. INLJ, SJ, and HJ), secondary index type (i.e. Eager, Lazy, and Comp), and consistency strategy (i.e. Synchronous (S) and Validation (V)). 
\revision{For example, INLJ-NS/S-Eager refers to the indexed nested loop join applied to a scenario where the data table R is a regular column and S is indexed using an eager index with a synchronous consistency strategy.}




\noindent\textbf{Datasets.}
We incorporate four real datasets and two synthetic datasets in our experiments as listed in Table ~\ref{tab:dataset}. These datasets encompass diverse typical workloads with distinct data distributions specified by duplication, matching rate, and skewness.
Specifically, duplication refers to the average number of replicas of primary keys ($c_r$, $c_s$) or join attributes ($d_r$, $d_s$) in the data table, matching rate ($\epsilon$) represents the ratio of the selected entries in a join, and skewness ($\theta$) reflects the shape of data distribution.

Following the existing works related to joins~\cite{marcus2021flow,wu2023factorjoin,wu2021unified,sun13end,sabek2023case,marcus2020benchmarking}, we examine three widely used benchmarks, Stack~\cite{marcus2020bao}, IMDB~\cite{leis2015good}, and SOSD~\cite{marcus2020benchmarking} to derive our real datasets. For tailoring each benchmark to our task, we select two joinable attribute as the join attributes.
{\it User} is derived from the Stack benchmark, which involves high duplication and skewness along with a moderate matching rate.
{\it Movie} dataset is extracted from IMDB benchmark that has high duplication and skewness while presenting a low matching rate.
{{\it Face} and {\it Wiki} dataset are selected from SOSD benchmark which presents high matching rate and slightly skewed distribution with a few duplications~\cite{marcus2020benchmarking}, respectively.}

Furthermore, we design two synthetic datasets that enable swift adjustment of features like entry size and skewness to generate diverse workloads. {\it Unif} dataset uses uniformly distributed integers as join attributes, while {\it Zipf} dataset uses widely utilized Zipfian distribution~\cite{huynh2024towards,lu2017wisckey,mo2023learning,huang2023design,schuh2016experimental,wu2013predicting} to generate skewed join attributes.

\textcolor{black}{By default, each dataset contains 10 million entries (64 bytes per entry, with 10-byte primary keys and join attributes), resulting in a total size of 0.5GB. This setup balances capturing key LSM-tree characteristics with experimental efficiency. We also vary dataset sizes from 0.5GB to 40GB in later experiments for a broader analysis.}

\vspace{-3mm}
\subsection{Current join methods can be suboptimal}
Current LSM-based stores typically adhere to conventions in relational databases that determine join algorithms based on selectivity and neglect many potentially effective index designs.
Our experiments indicate that several factors also significantly impact performance of different join methods, like entry size (Figure ~\ref{fig:joinsel} (a)), join frequency (Figure ~\ref{fig:joinsel} (b)), and matching rate (Figure~\ref{fig:joinsel} (c)). Furthermore, various index designs exhibit impressive performance under specific conditions, influenced by entry size (Figure~\ref{fig:5.7}), data distribution (Figure~\ref{fig:5.8_unif} and Figure ~\ref{fig:5.8_zipf}), and join frequency (Figure~\ref{fig:5.9}).
To the best of our knowledge, no existing works have thoroughly investigated the impact of these factors on the join methods selection or carefully examined the underlying principles. Our study fills this gap and provides novel insights from extensive experiments.

\vspace{-3mm}
\subsection{Entry size vs. join algorithm selection}
\label{subsec:exp_alg_entry_size}
\textcolor{black}{We test different join algorithms on two real datasets, {\it User} and {\it Face} that present different performance trends, with entry sizes varying from 32 bytes to 2048 bytes.} 
\textcolor{black}{We evaluate a wide range of scenarios, including those without index, single index, and dual index, as presented in Figure ~\ref{fig:joinsel} (a). In single index cases, we focus on configurations with a primary index to avoid the impact of index construction costs. For dual index scenarios, INLJ and HJ are excluded since, as Table ~\ref{tab: join_cost} indicates, additional indexes do not lead to performance enhancement. Moreover, those without index typically present inferior performance, thus being placed in our technical report for space constraints. Furthermore, Composite Index and synchronous strategy are employed by default, and the impact of index types is discussed in subsequent subsections.}

The results show that SJ and HJ have comparable join latencies across various cases and datasets, consistent with our theoretical analysis. The join costs of SJ-P and HJ-P are $(N_{S} \cdot \frac{e_S}{B}) + 5(N_R \cdot \frac{e_R}{B})$ and $3(N_S \cdot \frac{e_S}{B}) + 3(N_R \cdot \frac{e_R}{B})$, respectively, which are nearly identical when tables $R$ and $S$ have similar data scales and entry sizes. Both algorithms exhibit a linear increase in join cost as entry size grows, aligning with the experimental results. In contrast, INLJ shows greater resilience to larger entry sizes. For example, as entry size increases from 32 to 4096 bytes in the {\it Face} dataset, HJ’s join latency increases sixfold, while INLJ’s latency only rises by 50\%. As analyzed in Section ~\ref{subsec:join}, INLJ's overhead includes scanning and searching costs, the former grows with entry size, but the latter remains constant if the entry size is smaller than the block size. This explains why INLJ, initially performing worse with small entries, gradually outperforms SJ and HJ as the entry size increases. In the {\it User} dataset, INLJ's latency is 3 seconds higher than SJ and HJ with 32 byte entries but surpasses both when the entry size exceeds 512 bytes.

\revision{Notably, sort-merge join with dual index (SJ-PS) exhibits impressive join efficiency due to the efficient range scan.} However, it requires significant index building overhead, degrading the overall performance. Meanwhile, the impact of index building cost varies with join frequency which we explore in the next subsection.

\vspace{-2mm}
\subsection{Join frequency vs. join algorithm selection}
\label{subsec:exp_alg_join_freq}
To explore the impact of join frequency on algorithm selection, we vary join frequencies from 1 to 32 and evaluate the performance of various join algorithms.
\textcolor{black}{The dataset selection follows the criteria with Figure ~\ref{fig:joinsel} (a), with the inclusion of {\it Wiki} for diversity. Here, join frequency refers to the number of join operations in a fixed workload. For example, with a join frequency of 2, a dataset with 10 million updates is split into two subsets of 5 million updates each, and a join operation is performed after processing each subset. Thus the total number of updates remains constant across experiments, regardless of join frequency.} We assessed system latency including both join and index-building latency, as shown in Figure ~\ref{fig:joinsel} (b). 
The results show that SJ-PS performs poorly at low join frequencies but gradually surpasses other algorithms as frequency increases. \revision{For example, with a join frequency of 1, the system latency of SJ-PS is nearly five times that of the best-performing algorithm, hash join with primary index (HJ-P).} However, with join frequency increases to 32, SJ-PS becomes the best algorithm with system latency $50\%$ of HJ-P.
As detailed in Table ~\ref{tab: join_cost} and Figure ~\ref{fig:joinsel} (a), SJ-PS achieves high join efficiency because its indexes allow sequential retrieval by join attributes without sorting. This comes at the cost of extra index construction overhead. As join frequency rises, the latency of SJ-P, INLJ-P, and HJ-P increases substantially, while SJ-PS grows more moderately, making it more appealing for high-frequency joins.

INLJ-P typically has higher system latency than SJ-P and HJ-P for small entry sizes, as discussed in Section ~\ref{subsec:exp_alg_entry_size}. However, INLJ outperforms SJ in the {\it User} dataset, which challenges our expectations. Differences in matching rates, duplication, and skewness between the datasets highlight the significance of these factors in selecting join methods, which is analyzed in the following section.

\vspace{-3mm}
\subsection{Matching rate vs. join algorithm selection}
\label{subsec:exp_alg_match_rate}
As discussed in Section ~\ref{subsec:exp_alg_join_freq}, matching rate can significantly impact join algorithm performance, influencing the selection of the most suitable method. \textcolor{black}{Since matching rates in real datasets cannot be adjusted, we use the representative dataset {\it Unif} to evaluate various join algorithms, with matching rates decreasing from 1.0 to 0.2, where the join attribute is a non-primary key in both tables $R$ and $S$.}
We break down system latency into joining and index-building latency, as shown in Figure ~\ref{fig:joinsel} (c). It is observed that INLJ's join cost is positively correlated with the matching rate. This is because lower matching rates lead to fewer matched entries and more empty point lookups, which cost less than non-empty lookups, as indicated in Table ~\ref{tab: sec_index}. In contrast, HJ and SJ join latencies are unaffected by matching rate, as they must process all entries regardless of matches.
\textcolor{black}{This experiment also reveals the general impact of secondary indexes on the performance of different join algorithms. The update performance of secondary indexes affects the index building latency of INLJ and SJ, while the lookup performance mainly influences the joining latency of INLJ.} More complete and in-depth analysis is provided in the following subsections.

\vspace{-1mm}
\begin{tcolorbox}
\vspace{-1mm}
\noindent {\bf Remark:} 
{\bf Join algorithm selection is influenced by more than just selectivity}. Larger entry sizes or lower matching rates favor INLJ, while higher join frequency reduces the impact of index-building overhead, making index-based algorithms like INLJ and SJ more advantageous. When index-building costs are negligible, SJ with two indexes is the preferred choice.
\vspace{-1mm}
\end{tcolorbox}
\vspace{-2mm}

\begin{figure}[t!]
\centering
  \includegraphics[width=\linewidth]{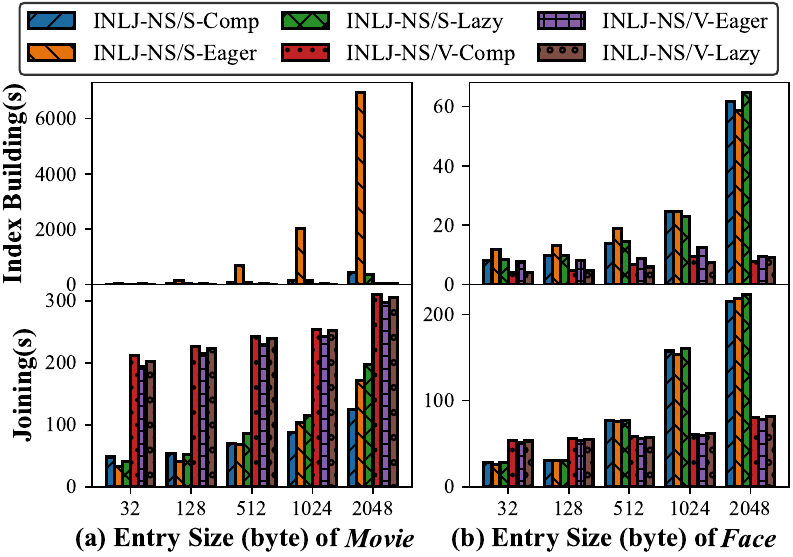}  
        \vspace{-4mm}
  \caption{The Impact of entry size on index designs.}
\label{fig:5.7}
    \vspace{-3mm}
\end{figure}

\begin{figure*}[t!]
\vspace{-3mm}
\centering
  \includegraphics[width=0.88\linewidth]{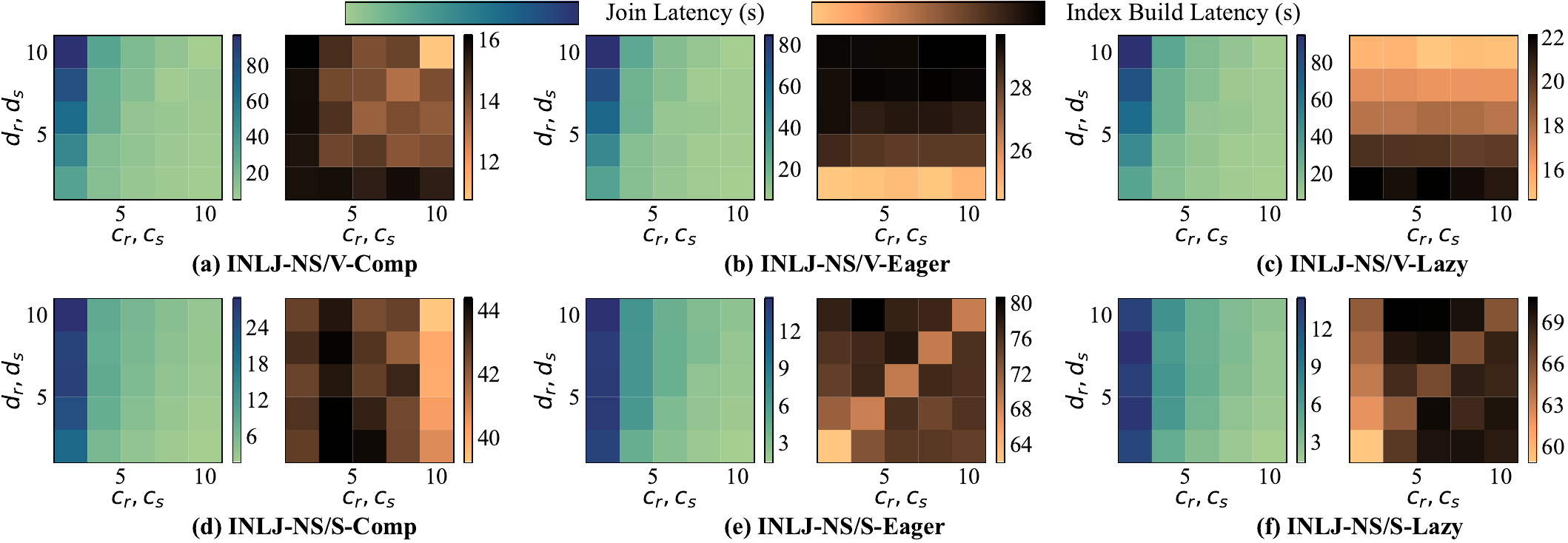} 
      \vspace{-3mm}
  \caption{The performance of various index designs with diverse primary key and join attribute duplication in {\it Unif} dataset.}
\label{fig:5.8_unif}
    \vspace{-2mm}
\end{figure*}

\subsection{Entry size vs. Index design}
\label{subsec:exp_index_entry_size}
\textcolor{black}{To examine the impact of entry size on index design, we evaluated different entry sizes on real datasets $Movie$ and $Face$, representing the more skewed cases and more uniform (less duplicated) scenarios, respectively.} 
For the Eager Index with Synchronous, the index building costs increase obviously with entry size on skewed dataset $Movie$, reaching 7000 seconds for 2048-byte entries as is shown in Figure~\ref{fig:5.7}. This is due to the exhaustive merging of entries with the same keys. Moreover, in skewed datasets, some entries occur more frequently, leading to dramatically larger entry sizes. Synchronous has to retrieve these large entries repeatedly to remove stale values. In this process, large entry sizes can lead to a more serious burden. This is due to the frequent merging of large, repeated entries, which significantly increases the overhead of retrieving and removing stale values. In contrast, Composite and Lazy Indexes, which don’t eagerly merge entries, handle entry size more efficiently, resulting in relatively stable index building costs.

In uniform datasets like $Face$, Synchronous index building costs remain stable due to fewer merges, but its join costs surpass those of Validation, since Validation avoids repeated checks in less-duplicated data. Moreover, Synchronous must maintain a large index table, with lookup overhead, as analyzed in Table~\ref{tab: sec_index}, increases with entry size. In contrast, Validation, storing only primary keys and join attributes, has more stable index sizes, making it preferable for large entries in uniform datasets. For skewed data, the Eager Index is inefficient due to high index building time, especially with large entries. The composite key index is more favorite given it generally delivers sound performance. However, this preference is also subject to the join frequency as we introduce in Section ~\ref{subsec:exp_index_join_freq}.

\vspace{-3mm}
\subsection{Data distribution vs. Index design}
\label{subsec:exp_index_distribution}
We examine data distribution's impact on index design in two scenarios. 
\textcolor{black}{Since data distribution is determined in real datasets, we include synthetic datasets to analyze the duplication in uniform datasets by varying the duplicates of primary key  ($c_r,c_s$) and join attribute in a grid pattern on {\it Unif} dataset, as well as investigate the impact of join attribute skewness ($\theta_r,\theta_s$) on {\it Zipf} dataset.}

\noindent {\bf Duplication.} As Figure~\ref{fig:5.8_unif} reveals, join latency decreases with increment of primary key duplicates, since LSM-tree merges identical primary keys that effectively reduces data scale. For all Validation methods (Figures~\ref{fig:5.8_unif}(a)-(c)), join costs increase with higher join attribute duplication. This aligns with our analysis in Table~\ref{tab: sec_index}, indicating the increased lookup cost of Validation. For Eager with Validation, the cost is represented as $O(d \cdot (L \cdot p + \lceil \frac{e}{B} \rceil)) + O(L' \cdot p + \lceil \frac{e'}{B}\rceil)$. As the duplication $d$ increases, so does the overall cost. In the Synchronous approach, only Composite Index shows a slight increase in join latency with higher join attribute duplication (Figure~\ref{fig:5.8_unif}(d)). This is because Eager Index and Lazy Index can merge more identical keys into single entries, consistent with our analysis.

Regarding index building, Eager Index demonstrates increased latency with higher join attribute duplication (Figure~\ref{fig:5.8_unif}(b)) due to non-empty point lookups for each duplicate. In contrast, Lazy Index shows an opposite trend (Figure~\ref{fig:5.8_unif}(c)) because the compaction process in LSM-tree merges identical join attributes, which not only avoids extra lookup costs but also reduces the index table size. For Synchronous methods, the trend is less clear. Higher primary key duplication leads to more entry checks but also a smaller index table, causing index building costs to fluctuate. Consequently, for Composite Index (Figure~\ref{fig:5.8_unif}(d)) and Lazy Index (Figure~\ref{fig:5.8_unif}(f)), index build costs remain relatively stable.

\begin{figure}[t]
\centering
  \includegraphics[width=\linewidth]{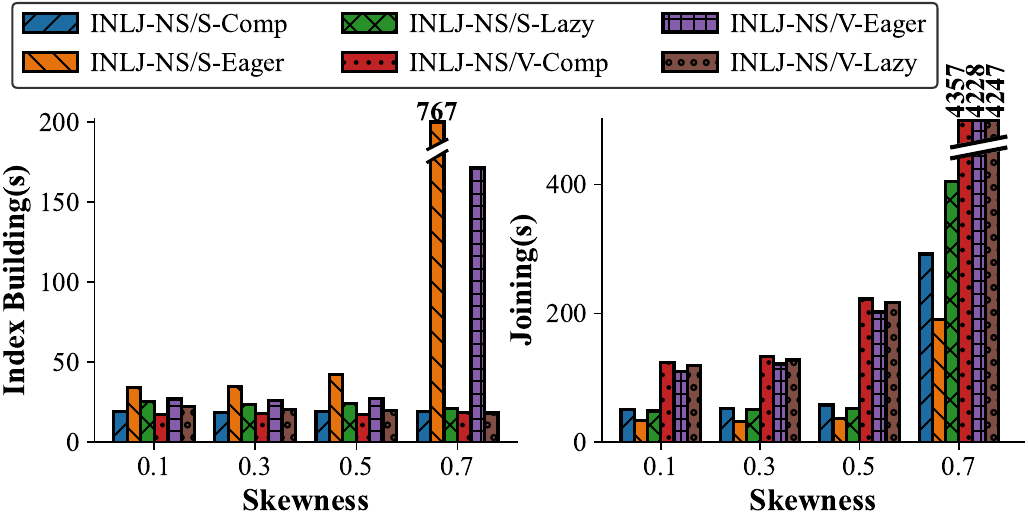}  
  \vspace{-4mm}
  \caption{The performance of various index designs with varying skewness in {\it Zipf} dataset.}
\label{fig:5.8_zipf}
\vspace{-3mm}
\end{figure}

\noindent {\bf Skewness.}  Figure~\ref{fig:5.8_zipf} reveals striking patterns in index building and join latency as data skewness increases. Notably, Eager Index demonstrates exceptionally high index building latency, while all Validation methods exhibit dramatically high join latency. \revision{When skewness reaches 0.7, the eager index enhanced INLJ with synchronous (INLJ-NS/S-Eager) and validation (INLJ) strategy require 767 and 172 seconds for index building, respectively, while other methods take less than 25 seconds.} This substantial difference can be attributed to the increased frequency of validity checks for high-occurrence entries as skewness grows. Furthermore, Eager Index's practice of merging identical keys results in potentially very large entry sizes for high-occurrence items. This issue is exacerbated in the Synchronous approach, which must frequently retrieve these large entries to remove stale primary keys from the value. In contrast, Lazy Index and Composite Index, which don't eagerly merge entries, maintain more manageable entry sizes.

On the join latency side, Validation methods incur significantly higher costs, which exceed 4000 seconds when skewness approaches 0.7, while other methods take only a few hundred seconds. This can be explained by our analysis in Table~\ref{tab: sec_index}, where each Validation method includes a lookup term of $O(d \cdot (L\cdot p + \lceil \frac{e}{B} \rceil))$. This term scales with increasing skewness and associated factor $d$, especially for INLJ, thus leading to the observed high join latencies. In summary, Lazy Index and Composite Index with Synchronous approach demonstrate superior resilience to data skewness.

\begin{figure}[t]
\centering
  \includegraphics[width=\linewidth]{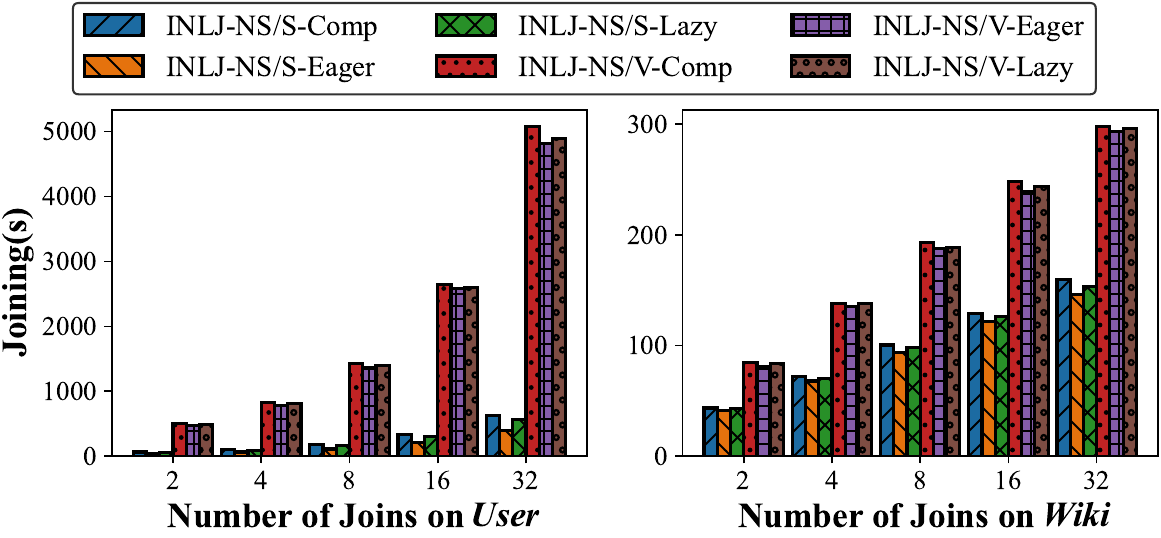}  
  \vspace{-5mm}
  \caption{Different index designs deliver varied performance facing increasing join frequency.}
\label{fig:5.9}
\vspace{-2mm}
\end{figure}

\begin{figure*}[t!]
\vspace{-3mm}
\centering
  \includegraphics[width=\linewidth]{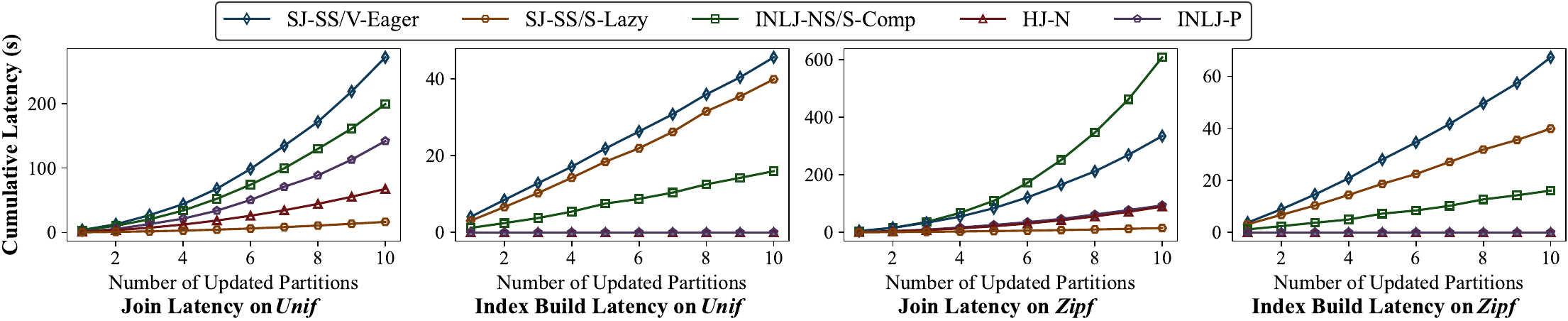}  
  \vspace{-6mm}
  \caption{The performance of different join methods under increasing database updates.}
\label{fig:update}
\vspace{-3mm}
\end{figure*}

\vspace{-3mm}
\subsection{Join frequency vs. Index design}
\label{subsec:exp_index_join_freq}
\textcolor{black}{We examine the impact of join frequency on index design by varying the join frequency on two datasets: {\it User} and {\it Wiki}, following the experiment in Section ~\ref{subsec:exp_alg_join_freq}.} 
As illustrated in Figure~\ref{fig:5.9}, the overall costs of Validation methods rise much faster than Synchronous methods as join frequency increases in both uniform ({\it Wiki}) and skewed ({\it User}) datasets. 
For example, in the {\it User} dataset, Validation costs jump from 460-500 seconds to 4800-5100 seconds, while Synchronous methods remain under 630 seconds. This difference stems from their design: Validation must check the data table for every matching entry, which results in repeated validity checks as join frequency increases. Synchronous methods, on the other hand, verify the validity of each entry only once during the index building phase, hence saving $O(d \cdot (L \cdot p + \lceil \frac{e}{B} \rceil))$ I/Os compared to Validation for each lookup in subsequent joins.

Moreover, Eager Index demonstrates the best overall performance among the three types of indexes. For instance, on the {\it User} dataset, Eager Index with Synchronous completes 32 joins in 390 seconds, while Composite and Lazy Index require 630 and 570 seconds. Eager Index saves 38\% and 32\% overall latency respectively, which will become more pronounced as join frequency increases. This advantage stems from Eager Index's lower join cost, which can be guaranteed as one entry lookup ($O(L' \cdot p + \lceil \frac{e'}{B} \rceil)$ as shown in Table~\ref{tab: sec_index}). In contrast, the other two index types may scatter entries with the same keys across different levels of the LSM-tree, necessitating multiple retrievals for a single matching data entry. The performance gap seems relatively modest with Validation methods as their costs are dominated by the check process of Validation. In summary, Synchronous and Eager Index methods demonstrate superior performance as join operations become more frequent. 

\vspace{-1mm}
\begin{tcolorbox}
\vspace{-1mm}
\noindent {\bf Remark:} 
{\bf Eager Index and Validation are underestimated in conventional LSM-tree databases}. Eager Index excels with frequent joins (e.g., more than twice per 10 million updates), while Validation outperforms Synchronous when entry sizes are large (e.g., over 512 bytes on the $Face$ dataset).
\vspace{-1mm}
\end{tcolorbox}

\vspace{-3mm}
\subsection{Impact of increasing database updates}
\label{subsec:exp_other_update}

Database updates affect multiple parameters with complex system overhead impacts, making theoretical analysis difficult. Here we measured cumulative join and index building latencies across different join algorithms, index types, and consistency strategies with varying update counts. In our experiments, we tested both {\it Unif} and {\it Zipf} datasets by dividing the dataset into equal subsets, performing joins after processing each subset, and recording cumulative latencies, which results are shown in Figure~\ref{fig:update}.

In uniform datasets, cumulative join latency grows quadratically, meaning join costs increase linearly with data scale. This is supported by Table ~\ref{tab: join_cost}, which shows that almost every component is linearly related to the number of entries. For skewed datasets, as the duplicates of join attributes increases, the joining cost for INLJ methods exceeds that of SJ, and the cumulative latency trend increases more rapidly compared to uniform datasets. This is due to the exponential growth in duplicate join attributes, which necessitates significantly more point lookups. However, when the skewed attribute is the primary key, the impact is reduced since the duplicates are automatically merged by LSM-tree, thus reducing the data scale and the number of required point lookups.

In terms of index building, the cumulative latency increases linearly in both datasets, indicating that the cost of incrementally updating the index with each partition of uniform data remains consistent. As analysed in Table ~\ref{tab: sec_index}, the number of levels is almost the sole dynamic factor in each term, and these levels do not change rapidly during index construction. So the expected cost of index building remains relatively stable as the data volume increases. 
Furthermore, the index building cost for the Eager Index also climbs more steeply. This is because the frequently occurring join attributes produce enlarged index entries, which requires more resources to access.

\vspace{-1mm}
\begin{tcolorbox}
\vspace{-1mm}
\noindent {\bf Remark:} In uniform datasets, join and index building costs grow steadily across methods, {\bf thus our selection criteria is still applicable}. For highly skewed datasets, we recommend HJ or SJ with Composite or Lazy Index to deal with large-scale updates.
\vspace{-1mm}
\end{tcolorbox}
\vspace{-1mm}

\vspace{-2mm}
\subsection{Impact of Tuning LSM-tree}
\label{subsec:lsm_exp}
\textcolor{black}{To better understand the interaction of LSM-tree and joins, we examine various LSM-tree parameters. We utilize the flexible synthetic dataset with uniform distribution, {\it Unif}, to explicitly reflect the impact of different LSM-tree structures.}

\begin{figure}[b]
\vspace{-2mm}
\centering
  \includegraphics[width=\linewidth]{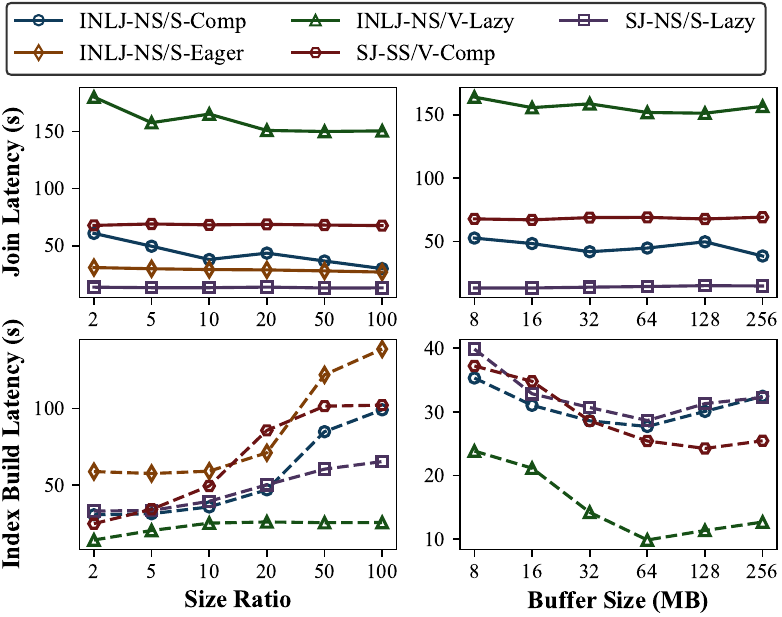} 
    \vspace{-8mm}
  \caption{The impact of size ratio and write buffer.}
\label{fig:lsm_para}
\vspace{-3mm}
\end{figure}

\noindent\textbf{Size ratio.}
Figure~\ref{fig:lsm_para} shows the impact of size ratio, a key compaction policy factor, which increases from 2 to 100. INLJ-NS/S-Comp join latency reduces from 60 to 30 seconds due to fewer LSM-tree levels and reduced Seek operations during lookups. However, Eager Index's join latency remains stable as it uses only one version of secondary keys. SJ and HJ are unaffected since lookup operations are not involved.
Meanwhile, index building latency increases with size ratio. For SJ-SS/V-Comp, it rises from 25 to 105 seconds as size ratio increases from 20 to 100, due to heavier update overhead caused by larger compactions. For INLJ, larger size ratios benefit frequent join scenarios, while smaller ratios are better for infrequent joins. For SJ and HJ, smaller size ratios are recommended to reduce index building cost and improve join performance.


\noindent\textbf{Write buffer.} 
Memory allocation has garnered significant interest within the LSM-tree community~\cite{mun2022lsm,huynh2021endure,luo2020breaking}. We examined memory allocation for the writer buffer, Bloom filters, and block cache, starting with the write buffer. In RocksDB, adjusting the write buffer size affects join latency for INLJ with Composite and Lazy Index by reducing levels and merging entries in index tables. This effect is less pronounced for Validation, where latency is predominantly influenced by the validation process. For SJ and HJ, as they do not rely on point lookups, changes in the number of levels do not affect their performance.
While larger write buffers theoretically reduce index building latency, they can lead to inefficient flushes and I/O spikes. Figure~\ref{fig:lsm_para} shows joining costs with 256MB buffer exceed those with 64MB buffer. We recommend using moderately large buffer sizes for better performance.

\begin{figure}[t!]
\centering
  \includegraphics[width=\linewidth]{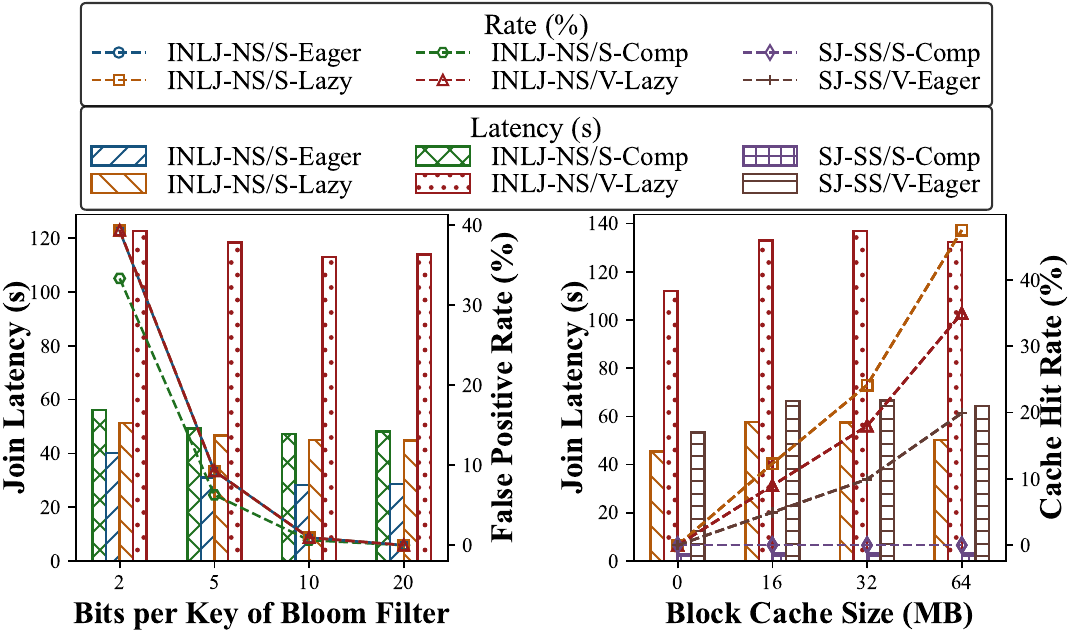}    
  \vspace{-5mm}
  \caption{The impact of block cache size and Bloom filter. }
    \label{fig:cache}
        \vspace{-1mm}
\end{figure}

\noindent\textbf{Bloom filter.}
Larger Bloom filter leads to reduced false positive rate that enhances INLJ method. As Figure \ref{fig:cache} shows, the join latency of INLJ-NS/S-Eager reduces by 30\% due to the drastically decreased false positive rate when we increase Bloom filter bits per key from 2 to 10. However, this improvement becomes marginal when we assign more memory, so we suggest using moderate memory (10 bits per key) to balance efficiency and memory usage.

\begin{figure*}[t]
\vspace{-5mm}
  \centering
  \hspace{-3mm}
  \raisebox{2mm}{
    \begin{subfigure}[b]{0.78\textwidth}
    \vspace{-20mm}
      \includegraphics[width=\linewidth]{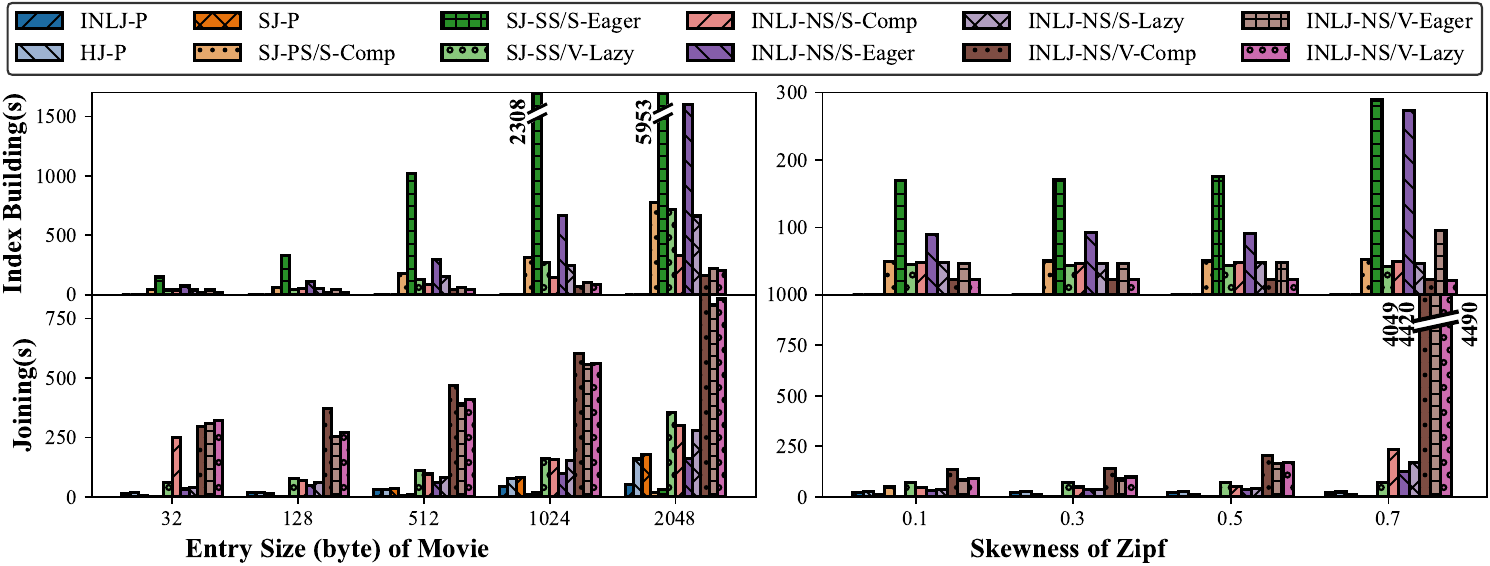}
      \vspace{-5mm}
      \caption{\textcolor{black}{Other LSM-based Systems: Pebble}}
      \label{fig:pebble}
      \vspace{-5mm}
    \end{subfigure}
    }
    \begin{minipage}[b]{0.20\textwidth}
      \begin{subfigure}[b]{\textwidth}
        \includegraphics[width=\linewidth]{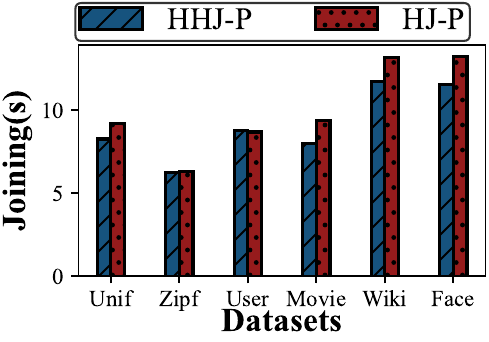}
        \vspace{-6mm}
        \caption{\textcolor{black}{Hybrid Hash Join}}
        \label{fig:hhj}
      \end{subfigure}
      \vspace{3mm}
      \begin{subfigure}[b]{\textwidth}
        \includegraphics[width=\linewidth]{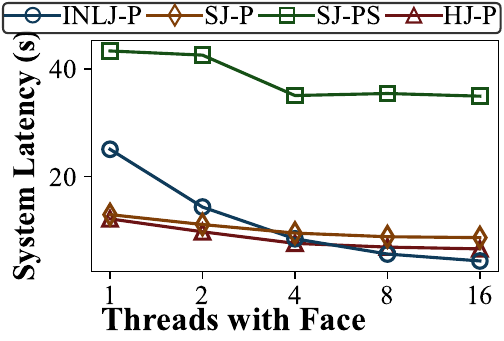}
        \vspace{-6mm}
        \caption{\textcolor{black}{Concurrency}}
        \label{fig:concurrent}
        \vspace{-5mm}
      \end{subfigure}
    \end{minipage}
    \caption{Case Studies of extending our insights to other system, hash join method, and concurrent execution.}
    \vspace{-2mm}
\label{fig:case}
\end{figure*}


\noindent\textbf{Block cache.}
While larger block cache should reduce join latency for INLJ and Validation methods, Figure~\ref{fig:cache} shows an obvious start up phase due to the additional CPU overhead caused by frequent updates when small caches are involved. SJ without Validation and HJ are not affected by cache size since they perform full data scans. Hence, we recommend using either default page cache or large block cache for INLJ and Validation for better performance.
\begin{tcolorbox}
\noindent {\bf Remark:} 
Larger size ratios or writer buffers improve INLJ's joining performance but raise index building costs. So we suggest moderate size ratios (10) and writer buffers (64-128MB). For INLJ, 10 bits per key is sound for Bloom filters. Block cache can be skipped if sufficient page cache is available.
\vspace{-1mm}
\end{tcolorbox}
\vspace{-1mm}

\section{Discussion}
\label{sec:highlight}
Here it is time to answer the question in title -- {\it are joins over LSM-trees ready?} The experiments and analysis in this benchmark indicate that there is still considerable potential for optimizing join method selection. The join methods used by existing databases cover only a small portion of our proposed configuration space and often fail to achieve optimal performance. Additionally, the efficiency of different joins varies significantly under many conditions. For instance, in Figure ~\ref{fig:joinsel} (b), SJ-PS could achieve more than 8x better latency than INLJ-P.
Fortunately, we have derived valuable insights that enhance our understanding of this topic and offer practical takeaways for developers in selecting appropriate join methods. Moreover, we point out some constraints of our benchmark and suggest future directions to guide development.
\vspace{-1mm}
\subsection{\textcolor{black}{Insights and takeaways}}
\label{subsec:discuss_insight}
\noindent{\bf One Method Does Not Fit All.} 
\revision{Figure ~\ref{fig:insight} summarizes the selection strategies for different join components, demonstrating that each method could excel under specific conditions, with no single method dominating universally. For instance, HJ-P shows competitive performance in many scenarios as illustrated in Figure ~\ref{fig:joinsel}. However, it is outperformed by SJ-PS with frequent joins and by INLJ-P when dealing with large entry sizes. 
This also suggests an important takeaway message for joins over LSM-tree: \textbf{\textit{Selection of join algorithms is influenced by multiple parameters (e.g., entry size, join frequency, and data distribution), instead of only subject to selectivity.}}}

\noindent{\bf Parameters matter unequally in join method selection.}
\textcolor{black}{Among the parameters examined, data distribution, join frequency, and entry size are the most influential ones. \textbf{\textit{Data distribution and join frequency are key for choosing secondary index types.}} Eager is preferred for frequent joins, excelling with skewed or highly duplicated join attributes, while Lazy and Comp offer lower index-building costs. \textbf{\textit{Consistency strategies depend on entry size and join frequency.}} Synchronous with a covering index is ideal for frequent joins with small entries, whereas Validation with a non-covering index works better otherwise. For skewed data, avoid pairing Synchronous with Eager or Validation with INLJ due to excessive overhead. Other parameters, like Bloom filters, size ratio, and block cache size, affect performance but don't significantly alter join method priority. Identifying these key factors simplifies the selection of join methods.}

\noindent {\bf Secondary indexes are not always necessary.}
Secondary indexes improve INLJ's join efficiency with Bloom filters and enhance SJ with reduced sorting costs. However, they come with certain index-building costs that sometimes can outweigh join improvements. As a result, secondary indexes may not always be beneficial, especially with large entries, high data skew, or infrequent joins.

\vspace{-2mm}
\subsection{Future Directions}
\label{subsec:future}
\revision{Multi-way joins are significantly more complex than binary joins that we discussed. Nevertheless, multi-way joins can be decomposed into a sequence of binary joins~\cite{marcus2018deep,marcus2019neo,marcus12plan,justen2024polar,marcus2021bao} thus our derived insights are still useful. For instance, we consider a join over three tables
$R \Join S \Join W$, where ($R \Join S$) aligns with our provided analysis and our insights still apply. This join produces an intermediate table ($V$) with $N_V$ tuples, which should be stored in an LSM-tree before joining with $W$, leading to an overhead $U_V$ of $O(L_V \cdot T \cdot N_V \cdot \frac{e_V}{B})$.
It's straightforward to make $V$ indexable to boost the subsequent join 
during LSM-tree construction. As a result, 
the costs of ($V \Join W$) for INLJ, SJ, and HJ are 
$U_V + (N_{W} \cdot \frac{e_{W}}{B}) + N_W((1 -\epsilon_{W}) Z_0(V) + \epsilon_{W} \cdot Z_1(V))$,
$U_V + (N_{V} \cdot \frac{e_V}{B}) + 5(N_W \cdot \frac{e_W}{B})$, and
$U_V + (3N_V \cdot \frac{e_V}{B}) + 3(N_W \cdot \frac{e_W}{B})$, respectively. 
This indicates that join method selection is heavily influenced by $N_V$. If $N_V$ is much larger than $N_W$, the costs pertaining to $W$ are negligible, indicating INLJ is preferable as it avoids extensive I/O overhead for reading the entire table $V$. If $N_V$ is comparable to $N_W$, the selection remains consistent with our previous analysis. For small $N_V$, INLJ is still preferable and SJ could outperform HJ after introducing a secondary index. Meanwhile, INLJ remains the best due to the desirable lookup cost achieved by the Bloom filter. These reveal enhanced attractiveness of INLJ in multi-way joins over LSM-tree. Moreover, adding a secondary index for subsequent binary joins becomes less critical, especially when $V$ is large, as an existing index is already sufficient in many scenarios.}
\revision{Additionally, many existing studies enhance join performance by optimizing binary join sequences to which our work could be adapted. Specifically, our theoretical models offer cost estimations and could be incorporated into optimization frameworks like Bao~\cite{khazaie2023sonicjoin} to generate optimized query plans by applying the above analysis to different join orders.. The specialized schemes and index structures like multi-way hash trie join ~\cite{khazaie2023sonicjoin} are beyond the scope of this paper.}

\revision{Moreover, LSM-based stores with different implementations and tuning parameters may present varied join performance thus yielding distinct insights, as our findings are based on experiments with RocksDB. However, we can easily extend our study to other platforms by adjusting configuration parameters and reevaluating the cost model for new insights. For example, platforms utilizing a key-value separation strategy like Wisckey ~\cite{lu2017wisckey,dai2020wisckey} offer higher update efficiency and may benefit from SJ due to faster index construction, which requires a reevaluation.} 
In such cases, our proposed configuration space can still guide the design of new evaluation frameworks, and many of our insights remain applicable since they reveal the fundamental principles of join operations over LSM-trees. 
\revision{Besides, we suggest several promising directions to improve joins over LSM-trees.
One is optimizing join algorithms specifically for LSM-trees and modern hardware~\cite{wang2023revisiting} to further boost performance. Another potential improvement lies in incorporating novel LSM-style~\cite{ferragina2020pgm} or hybrid secondary indexes~\cite{kim2020robust} to reach flexible update and query trade-off and enhanced overall performance. Moreover, advanced consistency strategies to maintain data integrity while reducing the query or update overhead are also crucial.
As data scales, distributed strategies and concurrent processing become vital for managing large datasets and high transaction volumes. Additionally, platform-specific optimizations, such as re-optimization schemes~\cite{kim2020robust} and normalization approaches~\cite{pebble}, could be further analyzed to identify the preferences for certain system.}

\subsection{\textcolor{black}{Extending to other LSM-based Storages}}
\label{subsec:case_study_pebbles}
\textcolor{black}{
We included another LSM-based system to validate the generality of the insights we provided.
Among these systems, Pebble ~\cite{pebble}, the core engine of CockroachDB~\cite{taft2020cockroachdb} written in Go, has garnered significant attention in the industry. Hence we use it to conduct a case study presented in Figure ~\ref{fig:case} (a).
As expected, although the exact performance on Pebble differ from those on RocksDB, the relative differences in the costs of different join methods, as well as the trends in performance changes across different settings, are apparently consistent with the results observed on RocksDB. Specifically, the joining latency for the HJ and SJ methods increases almost proportionally with the entry size, whereas the INLJ methods exhibit a slower increase. For instance, when selectivity is moderate, INLJ-P surpasses HJ-P and SJ-P in joining efficiency when the entry size reaches 512B. These observations are consistent with the phenomena we observed in Figure~\ref{fig:joinsel} (a). Additionally, in Figure ~\ref{fig:case} (b), the joining latency for the INLJ methods and Validation increases significantly under higher skewness, due to more frequent retrieval of higher-frequency data. When skewness reaches 0.7, the joining latency for all INLJ methods with Validation exceeds 4000 seconds, whereas other methods remain below 200 seconds. At the same time, the index building times for both Eager Index and Synchronize are noticeably longer than those of other methods. These results are also similar to the findings from the RocksDB experiments in Figure~\ref{fig:5.8_zipf}. Therefore, despite differences in implementation, since our insights are based on the theoretical cost model of LSM-tree, they can be generalized to other LSM-based systems.}

\textcolor{black}{
For the systems encompassing a subset of our configuration space, we can evaluate the associated space and utilize the corresponding insights or guidelines. In cases where join performance is altered by hybrid data structures or specific optimizations on certain join methods, we can adjust our findings to them with reevaluation since the instinct of each join method remains consistent.}


\vspace{-1mm}
\subsection{\textcolor{black}{Extending to other Hash Joins}}
\label{subsec:case_hash}
\textcolor{black}{ 
To further evaluate the generality of the insights presented in this paper across more specialized join methods, we also included hybrid hash join~\cite{dewitt1984implementation,shapiro1986join} as a case study. Compared to the default grace hash join used in this paper, hybrid hash join maintains an in-memory hash table for a portion of the table, potentially reducing I/O costs slightly. As shown in Figure ~\ref{fig:case} (b), across all datasets, hybrid hash join consistently exhibits a similar join cost to grace hash join, with a difference of only around one second. This is because joins over LSM-trees typically process massive data scale and performance is typically dependent on the I/O costs.
In this case, the memory budget is significantly smaller than the total data volume, meaning hybrid hash join can only store a limited portion of data in memory, thus offering only marginal optimization. Therefore, the derived insights are still applicable for hybrid hash join, which further indicates the generality of our work.}

\vspace{-2mm}
\subsection{\textcolor{black}{Supplementary analysis on Concurrency}}
\label{subsec:case_concurrency}
To further investigate the effects of potential variations in implementation, we examine the impact of concurrency in joins over LSM-trees on {\it Face}, a representative real dataset with uniform data distribution. Although concurrency does not appear to impact the I/O cost presented in Table ~\ref{tab: join_cost}, systems can still benefit from the parallel execution of join tasks which deserves further evaluation. To achieve this,
we divide the key ranges approximately evenly across threads, allowing each thread to scan a different range concurrently. These scans allow HJ to build hash tables in parallel, SJ to sort data of each range before merging them, and INLJ to perform concurrent point queries on the target data table. As shown in Figure~\ref{fig:case} (c), INLJ-P shows the best thread scaling, reducing latency by over 80\% at 16 threads due to minimal thread interdependence - each thread independently scans $R$R table and performs point lookups on $S$  table. SJ-P and HJ-P show less improvement as they are memory-bound, with threads waiting for memory resources. SJ-PS gains limited benefit as threading mainly affects index building compaction, which is heavily I/O-bound. Therefore, under memory constraints, INLJ methods demonstrate better concurrency potential.


\vspace{-2mm}
\section{Related Work}
\label{sec:related_work}
{\bf LSM-tree stores.} Extensive research has been conducted on LSM-trees, typically employing theoretical analysis to optimize parameters within a given design space, such as size ratio, compaction policy, and Bloom filters~\cite{dayan2018dostoevsky,dayan2017monkey,huynh2021endure,liu2024structural,mo2023learning,luo2020rosetta,dayan2022spooky,dayan2019log,huynh2023flexibility,yu2024camal,chen2024oasis,knorr2022proteus}. 
\textcolor{black}{Additionally, works focusing on self-designing data structures, like Cosine~\cite{chatterjee2021cosine}, Design Continuums~\cite{idreos2019design}, and Data Calculator~\cite{idreos2018data}, have been proposed to address specific challenges including varying data distribution, concurrency, and evolving hardware.
We agree that these methods could potentially improve join performance, while they typically enhance system performance by including more data structures like B+ tree in their configuration space instead of focusing on optimizing a certain type. Meanwhile, our results can also be utilized in these works to analyze the LSM component and optimize toward join operations since our insights are general for LSM-tree structure.}
Moreover, various LSM-based industrial data stores~\cite{rocksdb,google-leveldb,chang2008bigtable,cassandra,cao2022polardb, altwaijry2014asterixdb, taft2020cockroachdb} have supported join operation. While they typically provide several join methods for users to select by themselves. Thus, a comprehensive benchmark study of joins over LSM-trees is still lacking. The works focus on secondary indexes and consistency strategies~\cite{luo2019efficient,qader2018comparative,wang2023revisiting} are also related. 

\noindent{\bf Disk-based joins.} 
Disk-based joins differ from in-memory joins~\cite{huang2023design,sabek2023case,schuh2016experimental} primarily due to their larger data volumes, necessitating more disk I/O operations. While extensively studied in relational databases~\cite{jermaine2005disk,leis2015good,leis2018query,postgresql1996postgresql,MySQL,gaffney2022sqlite}, traditional wisdom suggests that join costs mainly depend on selectivity or cardinality~\cite{sun13end,hilprecht2019deepdb,absalyamov2018lightweight,wu2021unified,yang2020neurocard}. However, the LSM-tree context presents a different scenario. The unique structure of LSM-trees, compared to traditional storage engines like B+ trees~\cite{bayer1970organization,idreos2019design}, results in distinct write and read I/O costs, significantly impacting join performance.

\vspace{-1mm}
\section{Conclusion}
We provide a comprehensive analysis of LSM-tree join configurations and their theoretical foundations. Through extensive testing across various conditions and datasets, we derive several key LSM-specific insights and practical takeaways that could challenge current assumptions.

\begin{acks}
This research is supported by NTU-NAP startup grant (022029-00001). We thank all reviewers for their valuable suggestions.
\end{acks}
\bibliographystyle{ACM-Reference-Format}
\bibliography{sample}
\clearpage
\newpage
\mbox{}
\newpage
\end{document}